\documentclass[reqno,12pt]{article}
\usepackage{ae} % or {zefonts}0
\usepackage[T1]{fontenc}
\usepackage[ansinew]{inputenc}
\usepackage{amsmath}
\usepackage{amssymb}
\usepackage{graphicx}
\usepackage{color}
\usepackage[colorlinks]{hyperref}
\usepackage{epsfig}

\newcommand{\Q}{{\cal Q}}
\newcommand{\AAA}{{\cal A}}
\newcommand{\MM}{-_{\!\!-}}

\newcommand{\beq}{\begin{equation}}
\newcommand{\eeq}{\end{equation}}
\newcommand\beqa{\begin{eqnarray}}
\newcommand\eeqa{\end{eqnarray}}
\newcommand\bea{\begin{array}}
\newcommand\eea{\end{array}}

\newcommand{\bPhi}{\bar\Phi}
\newcommand{\im}{{\rm Im}\;}

\newcommand\IM{{\rm Im}\,}

\def\Xint#1{\mathchoice
{\XXint\displaystyle\textstyle{#1}}%
{\XXint\textstyle\scriptstyle{#1}}%
{\XXint\scriptstyle\scriptscriptstyle{#1}}%
{\XXint\scriptscriptstyle\scriptscriptstyle{#1}}%
\!\int}
\def\XXint#1#2#3{{\setbox0=\hbox{$#1{#2#3}{\int}$}
\vcenter{\hbox{$#2#3$}}\kern-.5\wd0}}

\def\dashint{\Xint-}

\def\ccw{{\hspace{-2.5mm}\unitlength 0.1in
\begin{picture}(1.00,1.00)(8.8,-11.50)
\special{pn 8}%
\special{pa 1120 1120}%
\special{pa 1100 1100}%
\special{pa 1080 1120}%
\special{fp}%
\end{picture}%
\hspace{-0mm}}}

\newcommand{\nn}{\nonumber}

\newcommand{\neqa}{\nonumber\end{eqnarray}}
\newcommand{\la}[1]{\label{#1}}

\newcommand{\eq}[1]{eq.(\ref{#1})}

\newcommand{\eqs}[2]{eqs.(\ref{#1},\ref{#2})}
\newcommand{\Eq}[1]{Eq.(\ref{#1})}

\newcommand{\QQ}{\mathfrak{Q}}

\newcommand{\half}{\frac{1}{2}}

\renewcommand{\d}{\partial}

\newcommand{\<}{{\langle}}
\renewcommand{\>}{{\rangle}}

\newcommand{\re}{\relax{\rm I\kern-.18em R}}

\def\su2{{SU(2)}}

\def\a{{\alpha}}

\def\[{\left[}
\def\]{\right]}

\def\s{\sigma}
\def\a{\alpha}
\def\b{\beta}
\def\th{\theta}

\def\[{\left[}
\def\]{\right]}

\def\<{\langle}
\def\>{\rangle}

\def\i2{\frac{i}{2}}

\def\bPhi{{\bar \Phi}}

\def\bQ{{\bar Q}}

\def\bPhi{{\bar \Phi}}

\usepackage{varioref}
\usepackage{makeidx}
\makeindex

\usepackage[english]{babel}
\begin{document}

%%%%%%%%%%%%%%%%%%%%%%%%%%%%%%%%%%%%%
%%%%%%%%%%%%%%%%%%%%%%%%%%%%%%%%%%%%%
%%%%%%%%%%%%%%%%%%%%%%%%%%%%%%%%%%%%%

\thispagestyle{empty}
\begin{flushright}\footnotesize
%\texttt{arxiv:yymm.nnnn}\\
\texttt{AEI-08-NNNN}\\
\texttt{LPTENS-08/68}\\
\texttt{DESY 08-193}\\
\vspace{1.7cm}
\end{flushright}

\renewcommand{\thefootnote}{\fnsymbol{footnote}}
\setcounter{footnote}{0}
\setcounter{figure}{0}
\begin{center}
{\Large\textbf{\mathversion{bold} Finite Volume Spectrum of 2D Field Theories\newline  from Hirota Dynamics}\par}

\vspace{1.0cm}

\textrm{Nikolay Gromov$^{\alpha}$, Vladimir Kazakov$^{\beta}$ and Pedro Vieira$^{\gamma}$}
\\ \vspace{1.2cm}
\footnotesize{
\textit{$^{\alpha}$
DESY Theory, Notkestr. 85 22603 Hamburg Germany;\\
II. Institut f\"{u}r Theoretische Physik Universit\"{a}t Hamburg
Luruper Chaussee 149 22761 Hamburg
Germany;\\
St.Petersburg INP, Gatchina, 188 300, St.Petersburg, Russia } \\
\texttt{nikgromov AT gmail.com}
\vspace{3mm}

\textit{$^{\beta}$ Laboratoire de Physique Th\'eorique
de l'Ecole Normale Sup\'erieure et l'Universit\'e Paris-VI,
75231, \\ Paris~CEDEX-05, France;} \\
\texttt{kazakov AT lpt.ens.fr}
 \vspace{3mm}

\textit{$^{\gamma}$ Max-Planck-Institut f\"ur Gravitationphysik, Albert-Einstein-Institut, Am M\"uhlenberg 1, 14476 Potsdam, Germany; \\ Departamento de F\'\i sica e Centro de F\'\i sica do
Porto Faculdade de Ci\^encias da Universidade do Porto Rua do Campo
Alegre, 687, \,4169-007 Porto, Portugal} \\
\texttt{pedrogvieira AT gmail.com}
\vspace{3mm}}

%%%%%%%%

\par\vspace{1.5cm}

\textbf{Abstract}\vspace{2mm}
\end{center}

\noindent
\small
We propose, using the example of the $O(4)$ sigma  model, a general method
for solving integrable two dimensional relativistic sigma models in
a finite size periodic box. Our starting point is the so-called
Y-system, which is equivalent to the thermodynamic Bethe ansatz equations of Yang
and Yang. It is derived from the Zamolodchikov scattering theory in the cross channel, for
 virtual particles  along the non-compact direction of the space-time
cylinder. The method is based on the integrable Hirota dynamics that follows from
the Y-system. The outcome  is a nonlinear integral equation for a
single complex function, valid for an {\it arbitrary quantum state}  and accompanied by the finite size analogue of Bethe
equations. It is close in spirit to the Destri-deVega (DdV) equation.
We present the numerical data for the energy of various states as
a function of the size, and derive the general L\"uscher-type formulas for
the finite size corrections. We also re-derive by our method the DdV
equation for  the $SU(2)$ chiral Gross-Neveu model.

\vspace*{\fill}

\setcounter{page}{1}
\renewcommand{\thefootnote}{\arabic{footnote}}
\setcounter{footnote}{0}

\newpage

%%%%%%%%%%%%%%%%%%%%%%%%%%%%%%%%%%%%%
%%%%%%%%%%%%%%%%%%%%%%%%%%%%%%%%%%%%%
%%%%%%%%%%%%%%%%%%%%%%%%%%%%%%%%%%%%%

{\footnotesize
\tableofcontents
}

\newpage
%%%%%%%%%%%%%%%%%%%%%%%%%%%%%%%%%%%%%
%%%%%%%%%%%%%%%%%%%%%%%%%%%%%%%%%%%%%
%%%%%%%%%%%%%%%%%%%%%%%%%%%%%%%%%%%%%

\section{Introduction and Summary}

The study of the properties of Quantum Field Theories (QFT's) in finite volume, or at finite
temperature, has a long history and numerous applications. Matsubara
description \cite{Matsubara:1955ws} of finite temperature $T$
thermodynamics, by considering the system in the periodic imaginary
time $t$, has lead to the extensive study of the Euclidean QFT's with
one compactified dimension with numerous physical applications
\cite{AbriGorTzya}.

 L\"uscher
found the leading finite size corrections to the mass gap in
relativistic two dimensional QFT's \cite{Luscher:1986pf,Luscher:1985dn}. These corrections depend solely on the asymptotic S-matrix of the theory.  Recently,   L\"uscher
corrections to various
multi-particle states in integrable 2D QFT  were conjectured  \cite{BJ}.

For the integrable 2D QFT's, as understood during the last two decades, the ambitions can be much higher: these systems are usually
solvable at any finite size though a systematic approach to such solutions, as well as  a good understanding of the working prescriptions, are still missing.

\begin{figure}[ht]
\epsfxsize=6cm
\centerline{\epsfbox{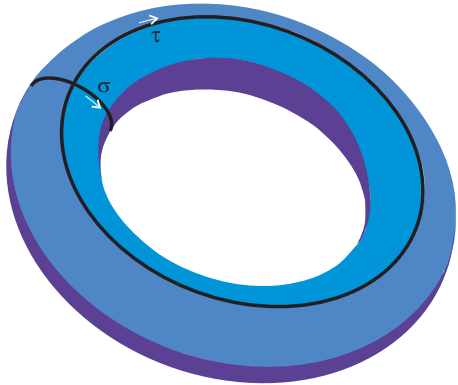}}
\caption{\label{pic2}
\small\small\textsl{Physical channel, cross-channel and finite volume vs finite temperature.
} }
\end{figure}

\begin{figure}[t]
\epsfxsize=15cm
\centerline{\epsfbox{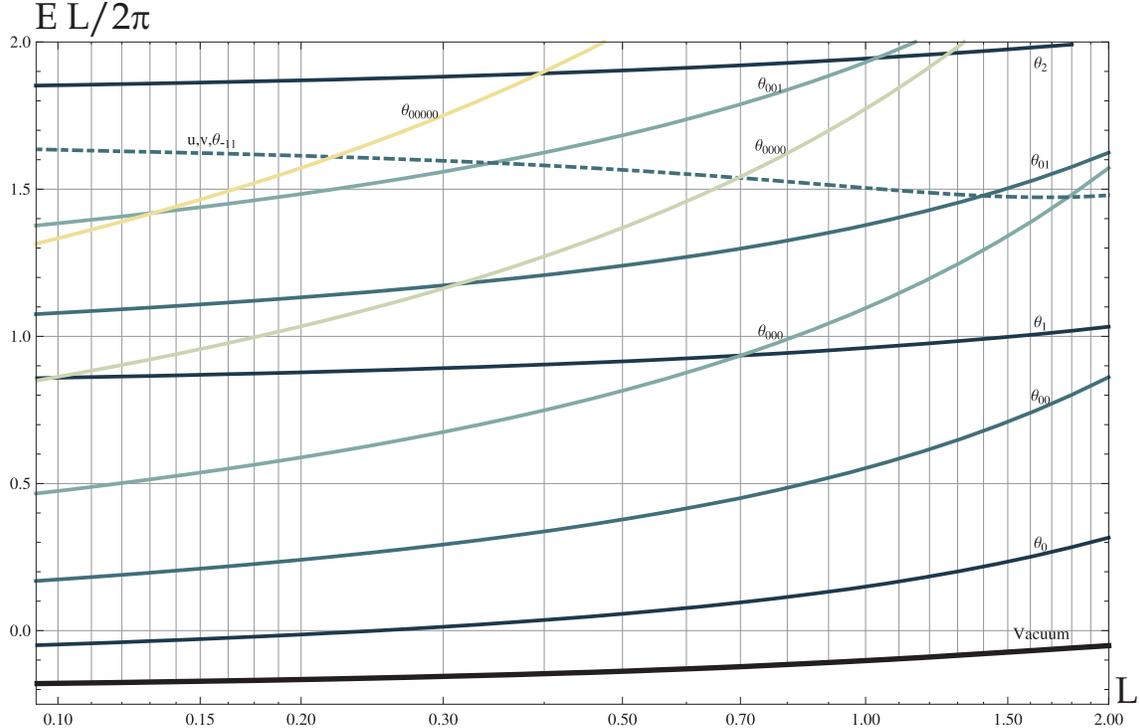}}
\caption{\label{pic4}
\small\textsl{ Plots of energies \(E\) of a few excited states  of $O(4)$ model on a circle of a circumference \(L\).
The vertical axis corresponds to the values of \(\frac{L}{2\pi}E\), the horizontal axis - to the values of \(L\) in the logarithmic scale. %The vertical axis corresponds to the values of \(\frac{E(L)L}{2\pi}\) and the vertical axis corresponds to  \(L \).
The lowest curve depicts the vacuum energy. The next one, labeled as
\(\theta_{0}\),  shows the mass gap energy. The corresponding state is
in the \(U(1)\) sector, with a single particle at rest, hence with the
mode number \(=0\).  The next states in the   \(U(1)\) sector are
denoted by \(\theta_{n_{1}n_{2}n_{3},\cdots }\), according to the  mode
numbers \(n_{1},n_{2},n_{3},\dots\) excited for the 1-st, 2-nd, 3-rd,
etc., particles. For all these states the $SU(2)_L$ and $SU(2)_R$ spins
of the several particles are pointing in the same direction, say they
are spin ``up". The dashed line represents a state having a polarization
out of the \(U(1)\) sector, with left and right ``magnons" excited - it
corresponds to the quantum state of two particles where both $SU(2)_L$
and $SU(2)_R$ spins are in the singlet $s=0$ state. The qualitative
explanation of these graphs will be given in subsection \ref{NUMexpl}.
} }
\end{figure}

\begin{figure}[t]
\epsfxsize=8cm
\centerline{\epsfbox{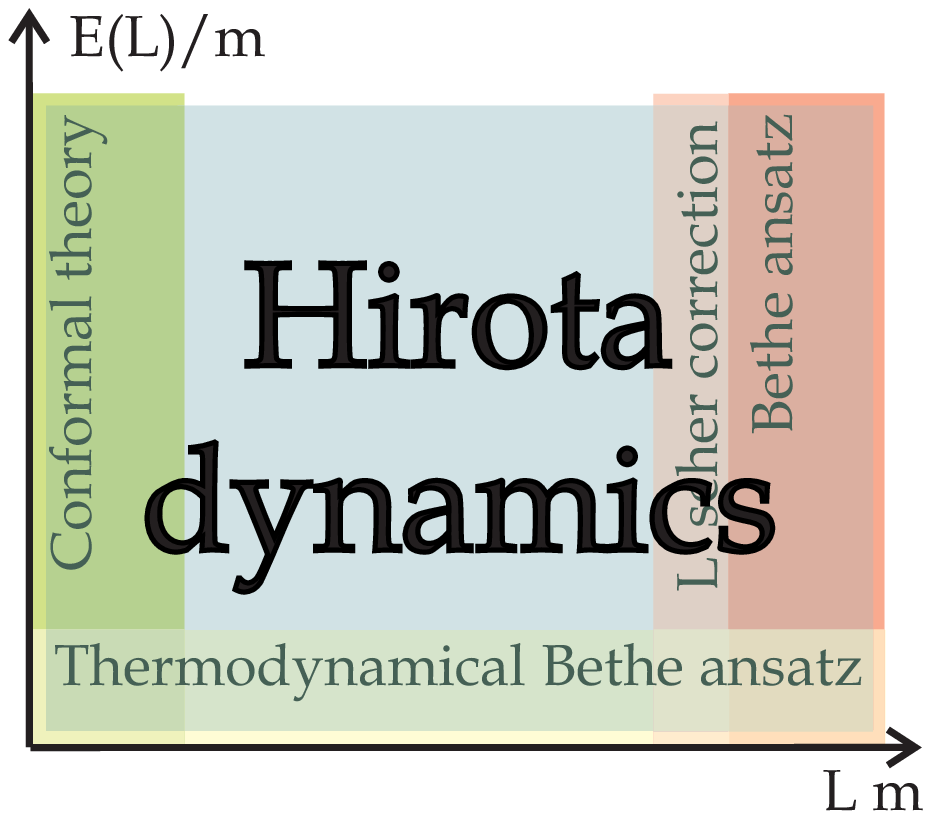}}
\caption{\label{pic1}
\small\textsl{Domains of applicability of different descriptions of an integrable field theory at a finite volume \(L\).
In the ultra-violet regime, for small volume measured in units of a
dynamically generated mass, the theory could be described by a
conformal theory. In the infrared, at large volume, one can use the
asymptotic Bethe equations. The leading order finite size corrections
are governed by the (generalized) L\"uscher corrections.  At any volume
but for the ground state energy only one can use Thermodynamical Bethe
ansatz. Hirota equation, equivalent to Y-system but more efficient when
it comes to imposing appropriate analyticity properties, is a universal
tool covering the whole diagram.}
}
\end{figure}

There are two main schemes to address the finite size calculations. The
first, pioneered by Destry and deVega (DdV) \cite{Destri:1987ug}, is
based on the integrable discretization. Once
such discretization is at hand, the system can be studied by the well
established methods based on the transfer matrix approach and the resulting non-linear integral equation (NLIE), often called the  DdV
equation, calculates not only the ground state energy but also the
spectrum of excited states. The method appeared to be very powerful
when applied to the Sine-Gordon model
\cite{Fioravanti:1996rz,Feverati:2000xa,Hegedus:2006st,Feverati:1999sr,Balog:2003xd},
or to more general RSOS models \cite{Zhou:1995hg},  Toda theories
\cite{ZinnJustin:1997at}, hard hexagon models
\cite{KluemperPierce:1991}, etc.

However, for generic integrable QFT it is far from easy to find the
corresponding integrable lattice regularization and for many models
such discretization is not known. Nevertheless, the problem can be
usually tackled by using a  computation scheme alternative to the DdV
 approach. As explained in the seminal  work of
Al.Zamolodchikov \cite{Zamolodchikov:TBA1990} this is achieved  by the
double Wick rotation trick: using the Matsubara imaginary time
formulation we can first find the free energy in the infinite volume
but finite temperature. Next we flip the meaning of euclidian time and
space directions on the cylinder: $\tau\to
\sigma,\,\, \sigma\to \tau$, and interpret the free energy as the ground state of the system in
finite volume $L=\frac{1}{T}$ (see fig.\ref{pic2}). In this way we can
obtain the exact finite volume ground state energy. This computational
scheme is known by the name of  Thermodynamic Bethe Ansatz (TBA).

The TBA equations, whose  number is infinite in many interesting models,  can  be usually concisely casted
into the so called Y-system functional equations
\cite{Zamolodchikov:1991vg,Kuniba:1993cn}. Often the latter one can be rewritten in the form of DdV
equations or some similar set of integral equations for a finite set of
functions. The method was successfully used for many relativistic
models \cite{Zamolodchikov:1991et,Martins:1991hi,
Fendley:1991xn,Fateev:1991bv,Bazhanov:1994ft}. As explained in the
previous paragraph the computation of the exact ground state energy by
means of this method is a relatively straightforward  task with
solid theoretical foundations.

To obtain the exact spectrum comprising all excited states of the
theory is, on the other hand, a much more involved - and a very interesting -
task.  A possibility to describe the excited states within the TBA approach, by modifying the analytical properties of the thermodynamic functions, was first suggested in \cite{Bazhanov:1994ft}. Another possible way to obtain the  spectrum of the theory, proposed around the same time, is based
on the analytic continuation of the ground state energy with respect
to the parameters of the model, such as the mass or the chemical
potential, in order to find the excited states \cite{Dorey:1996re}. If
the integrable lattice regularization is absent, it is not well
understood why these methods work. Nevertheless, the results are
usually in the excellent agreement with the perturbation theory,
L\"uscher finite size corrections and the direct Monte-Carlo study for
a wide range of sizes $L$ (see for example
\cite{Balog:2003yr,Balog:2005yz,Balog:2001sr,Hegedus:2005bg,Hegedus:2004xd}
for \(O(n)\) and related $\sigma$-models).

%Not only the ground state energy appears to be calculable but also
%sometimes the mass gap and a few other excited states.

For models with  diagonal scattering, like the Sinh-Gordon theory
\cite{Zamolodchikov:2000kt},  the whole classification of  excited  states is
possible \cite{Teschner:2007ng}. The situation is much more complex
when we deal with the non-diagonal scattering. The nested structure of the corresponding
Bethe ansatz equations leads to  complicated magnon-type excitations
and bound states. Little is known about the excited states in such finite size
systems. The only models where  the polarized excited states were
investigated, using the DdV equations, are the Sine-Gordon
model \cite{Fioravanti:1996rz} and its supersymmetric version \cite{Hegedus:2006st} as well as the tricritical Ising model \cite{Pearce:2000dv}. By the existing methods only the
sectors with diagonal scattering can be studied efficiently, as was
done for example for the \(O(4)\) sigma model in \cite{Hegedus:2004xd}.
A general and unified description of all  excited states of the
\(\sigma\)-models  like \(O(n)\) or the \(SU(N)  \)
principal chiral field (PCF), and similar ones, having a ``geometric"
target space, is still absent.

The main goal of the present paper is to give a method of  a general
and systematic description of all the states of integrable QFT's in
finite volume. We will explain how to go beyond the asymptotic spectrum
and compute the full finite size spectrum comprising all excited states
of integrable sigma models.
 We do it here on the example of $O(4)$ sigma model and also for the $SU(2)$ chiral Gross-Neveu model but our formalism is certainly more general and is most probably applicable to any integrable 1+1 dimensional \(\s\)-models. The main ingredients of the method are:
\begin{itemize}
\item  The two-particles  S-matrix for integrable system allows us to write the periodicity condition quantizing the momenta of the physical particles on a large circle of length \(R\). The equations following from the periodicity condition are  so called asymptotic Bethe ansatz (ABA) equations describing all states of the model. The details of this computation for the $SO(4)$ sigma model are given in Appendix A\footnote{We are unaware of the  existence of such derivation of the Y-system for the PCF in the literature}. They are, however, valid only in a sufficiently big volume compared to the typical interaction distance,  $Rm\gg 1$ where $m$ is the infinite volume mass gap.
\item  For the ground state, the double Wick rotation $(\sigma,\tau)\to(\tau,\sigma)$ allows to reduce the problem to the thermodynamics. One can put the euclidian theory on the torus with one radius, \(R,\) very large and another one, \(L\), arbitrary (see the fig.\ref{pic2}).
    The ground state energy for a finite radius is related to the thermodynamic partition function. The exact equations for it can be found using the asymptotic spectrum given in the cross channel by the asymptotic  Bethe equations. The resulting infinite series of integral equations,  thermodynamic Bethe ansatz (TBA) equations, are casted into a functional form  called Y-system. Here is the main assumption: we assume that  different solutions of the Y-system  describe not only the ground state but \textit{all} the excited states. One should furthermore restrict the class of solutions by assuming certain analytic properties which will in particular identify the quantum numbers of the states we are considering.
\item Classical integrability of the Y-system,  as a finite difference equation equivalent to the Hirota difference equation \cite{Hirota,Kuniba:1993cn}, allows us to express explicitly the infinite number of the unknown functions through a finite number of the basic ones \cite{Krichever:1996qd,Bazhanov:1998dq}.%\item A part of missing information for this  basic set of quantities is obtained from the analyticity %w.r.t. the spectral parameter (rapidity).
\item The Baker-Akhiezer function of the Lax pair associated with the Hirota equation can be interpreted as the Baxter function encoding the ``magnon" Bethe roots, responsible for the \(SU(2)_R\) and \(SU(2)_L\) polarizations of states. The analyticity properties important for the full formulation of the resulting non-linear integral equation, are also suggested by the Lax equations. The gauge symmetry of Hirota equations allows to explicitly fix the final nonlinear integral equation (NLIE) for each state of the theory.
\end{itemize}

The resulting  equation can be studied in various limits (such as  L\"uscher finite size corrections or small volume, conformal limit) or solved numerically in a rather efficient way.
The fig.\ref{pic4} shows some of our numerical results obtained from
the new equation, plotting the energy of various states as functions of
the volume. When the similar results are available in the literature
the agrement is perfect.

The general scheme elaborated in this paper on the example of the
$O(4)$ sigma-model should be applicable to all integrable relativistically
symmetric 2D QFT's. It should be also useful for the study of finite
size effects when the system does not look explicitly relativistic but
allows the S-matrix description and this S-matrix obeys the crossing
symmetry, like the AdS/CFT S-matrix \cite{Beisert:2006ez,Janik:2006dc}.
Y-system and Hirota equations give a  unified and powerful point of
view at all this subject since they solve in an almost trivial way the
``kinematic" part of the problem related to the representation theory, whatever is the symmetry or
supersymmetry of the model \cite{Krichever:1996qd,Kazakov:2007fy}. 
%The analytic properties of the
%functions entering Hirota equation (T-system) are also greatly
%constrained by its integrable structure.

Our method based on Hirota equation, being exact for any finite size
\(L\) of the system, reproduces well various limiting cases (see the
fig.\ref{pic1}). For the large \(L\), the energies of the states are
well described by the L\"uscher corrections
\cite{Luscher:1986pf,Luscher:1985dn,BJ}\footnote{Actually, as we will
see from our numerics,  L\"uscher corrections work surprisingly well
all the way until \(Lm\sim1\).}. We derive them here for a general
state with arbitrary polarization, which is also a new result,
extending some hypothesis existing in the literature \cite{BJ}. For
small \(L\), our results are well described by the theory of three free
bosons, as will be discussed in the paper. The results for various
low-lying levels, including the cases of non-diagonal scattering which
are new,  are summarized in the fig.\ref{pic4}.

Our resulting NLIE can be brought sometimes to a form similar to the
DdV equation. In the cases when the latter is available it can even coincide with DdV equation (an example of
the chiral Gross-Neveu model is considered in our paper). It would be extremely interesting to understand the relation between the solution based on the integrable lattice discretization of \cite{FR} and our proposed integral equations. Nevertheless we should stress that the real power of our method should be in its universality: it should work in all
situations when the TBA equations in the form of the Y-system are available.

%%%%%%%%%%%%%%%%%%%%%%%%%%%%%%%%%%%%%%%%%%

\section{TBA and Y-system for $O(4)$ sigma model, or  $SU(2)$ Principal Chiral Field}

The method we are proposing it quite general and we hope that a wide
range of models could be solved using it. However for the sake of
simplicity we will exemplify it on the $SU(2)$ Principal Chiral
Field (PCF), equivalent to the \(O(4)\) sigma model. In section \ref{sec:GN} we will also consider the $SU(2)$
Chiral Gross-Neveu model.
\subsection{The Model}

The action of the PCF is given by the usual expression
\begin{equation}
\mathcal{S_{\sigma}}_\text{} =\frac{1}{ e_{0}^{2} } \int dt \,dx \:\ (\partial_{\alpha} X_{a}^{_{}})^{2},\qquad\qquad\sum_{a=1}^{4}(X_{a}^{^{_{}}})^{2}=1\,,
\end{equation}
whose target space is $S^{3}$. It is equivalent to the $SU(2) \otimes
SU(2)$ principal chiral field (PCF) whose infinite volume solution
was given in \cite{Polyakov:1983tt,Wiegmann:PCF,Wiegmann:1984ec}.
Indeed, by packing the fields $X_i$ into an $SU(2)$ group element
$h=X_4+i\sum_{j=1}^3 X_j \sigma_j$ with $\sigma_j$ being the usual
Pauli matrices, we can re-write the action as\footnote{In the $AdS/CFT$ literature one
usually uses $\sqrt\lambda=\frac{4\pi}{e_0^2}$.}
% \begin{equation}
% g=\begin{pmatrix}X_{1}+iX_2 & -X_{3}+iX_4 \\
% X_{3}-iX_4 & X_{1}-iX_2 \\
% \end{pmatrix}\in SU(2)\\ \\
% \end{equation}
% leads to the action
\begin{equation}
\mathcal{\mathcal{S_{\text{PCF}}}}=-\frac{1{}}{2e_{0}^{2} } \int dt \,dx \:\ {\rm tr}(h^{-1}\partial_{\alpha} h_{}^{_{}})^{2}\,.
\end{equation}

The spectrum of this asymptotically free  theory in the infinite volume
consists of a single physical particle of mass $m= \Lambda
e^{-\frac{2\pi}{e_{0}^{2}}}$,  where $\Lambda$ is a cut-off. Its wave
function transforms in the fundamental representation under each of the  $SU(2)$
subgroups. Al. and A.Zamolodchikov \cite{Zamolodchikov:1978xm} proposed
the exact elastic scattering matrix for such particles:
\begin{equation}\label{eq:Smatr}
\hat {S}_{12}(\theta) =S_0(\theta) \frac{\hat R(\theta)}{\theta-i} \otimes S_0(\theta)  \frac{\hat R(\theta)}{\theta-i}\,, \qquad S_0(\theta)=i\frac{ \Gamma \left(\frac{1}{2}-\frac{i \theta}{2}\right) \Gamma
   \left(+\frac{i \theta}{2}\right)}{\Gamma \left(\frac{1}{2}+\frac{i
   \theta}{2}\right) \Gamma \left(-\frac{i \theta}{2}\right)}\;,
\end{equation}
where $\hat R(\theta)$ is the usual $SU(2)$ R-matrix in the fundamental
representation given by $
\hat R(\theta)=\theta+i \hat P
$ and $\hat P$ is the permutation operator exchanging the spins of the
particles being scattered. This S-matrix was established due to: (i)
analyticity, (ii) unitarity, (iii) absence of bound states, (iv)
crossing.  In particular, (ii) and (iv) lead to the following
identity
\begin{equation}
S_0(\theta+i/2)S_0(\theta-i/2)=\frac{\theta-i/2}{\theta+i/2}
\la{crossing}
\end{equation}
on the scalar (dressing) factor. We can use this S-matrix to study the spectrum
of $N$ particles in a periodic space circle of a sufficiently big
circumference $L\gg m^{-1}$. The spectrum can be defined from the wave function
periodicity condition
\begin{equation}
\prod^{N}_{j=k+1 }{\cal \hat S}(\theta_k-\theta_j)
\prod^{k-1}_{j=1 }{\cal \hat S}(\theta_k-\theta_j) |\Psi\rangle=e^{-imL\sinh(\pi\theta_k)}|\Psi\rangle\;, \label{perintro}
\end{equation}
which quantizes the momenta of the physical particles. The asymptotic
spectrum of the theory put on a large circle of length $L$ is then
given by
 \begin{equation}
 E=\sum_{j=1}^N m \cosh(\pi\theta_j)\, \label{energy}
 \end{equation}
where $\theta_j$ are solutions to the Bethe equation (see Appendix
\ref{AppA} for more details). In what follows we will measure all
dimensional quantities in the units of \(m\). Diagonalizing the periodicity condition
\eqref{perintro} in the physical space by the usual methods (see an appendix in \cite{Kazakov:2007fy} for this model) we get the main Bethe equation
\beqa
e^{-iL\sinh(\pi\theta_j)} = -\prod_k S_0^2(\theta_j-\theta_k)
\frac{\Q_u(\theta_j+i/2)}{\Q_u(\theta_j-i/2)}
\frac{\Q_v(\theta_j+i/2)}{\Q_v(\theta_j-i/2)} \,. \label{BAEmiddle0}
\eeqa
The magnon rapidities $u_j$ and $v_j$ are fixed by the auxiliary Bethe
equations
\beqa
-\frac{\Q_u(u_j+i)}{\Q_u(u_j-i)} =\frac{\phi(u_j+i/2)}{\phi(u_j-i/2)}   \,\, , \,\, -\frac{\Q_v(v_j+i)}{\Q_v(v_j-i)} =\frac{\phi(v_j+i/2)}{\phi(v_j-i/2)} \label{BAEuv0}\;,
\eeqa
where
\begin{equation}\label{PolyQF}
\Q_w(x)=\prod_j(x-w_j)\,,\qquad \text{for}\quad w=u,v,
\end{equation}
and $\phi(x)=\prod_{j}(x-\theta_j)$.

\subsection{TBA and Y-system}

As we mentioned in the introduction, the ground state energy $E_0(L)$ for
arbitrary $L$ can be computed starting from asymptotical Bethe ansatz
in the cross-channel. For  $SU(2)$ principal chiral field this is
described in detail in the Appendix
\ref{AppA}. The output is that the ground state energy is given by
\begin{equation}\label{E0:eq}
E_0(L)=-\frac{1}{2}\int d\theta\cosh(\pi\theta)\log(1+ Y_0)\;,
\end{equation}
where $Y_0$ is one out of an infinite number of $Y$-functions $Y_n$
with $n\in  \mathbb{Z}$ obeying the TBA-type equations
\begin{equation}\la{Y:eq0}
\log Y_n+ L\cosh(\pi x)\delta_{n0}=s*\log(1+Y_{n+1})(1+Y_{n-1}) \,,\qquad n=0\pm1,\pm2,\dots
\end{equation}
with $s=\frac{1}{2\cosh(\pi x)}$ and the sign $*$  denoting the convolution.
If $\log(Y_n(x))$ for any $n$ have no singularities inside the physical strip
$-1/2<\im x<1/2$ we can easily invert the operator $s*$ to get simply
$s^{-1}=e^{\frac{i}2{\partial _x}}+e^{-\frac{i}2{\partial _x}} $ and
these integral equations can be rewritten in a  functional, $Y$-system
form
\begin{equation}\la{Y:eq}
Y^+_nY^-_n=(1+Y_{n+1})(1+Y_{n-1}) \,,
\end{equation}
supplemented with the asymptotic boundary conditions for large $x$
\begin{equation}
Y_n\sim e^{- L\cosh(\pi x) \delta_{n0}} \,\times\text{const}_{n}\,.
\label{asympt}
\end{equation}
The superscripts $\pm$ stand for shifts of the argument by $\pm i/2$ \footnote{We will often use even a more general notation, like \(f^{\overbrace{++\ldots+}^{k} }=f(x+ik/2)  \) or \(f^{\overbrace{--\ldots-}^{k}\ }=f(x-ik/2)  \).},
\begin{equation}
 f^{\pm}\equiv f(x\pm i/2) \,.
\end{equation}
%%%%%%%%%%%%%%%%%%%%%%
\begin{figure}[h]
\epsfxsize=5cm
\centerline{\epsfbox{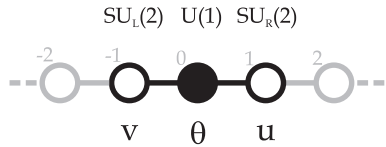}}
\caption{\label{pic3}
\small\textsl{Dynkin diagram (three central nodes) and its extension for the magnon bound states (grey nodes) reflecting the structure of the Y-system. The central, black node corresponds to the U(1) sector excitations of the model (\(\th\)-roots), the upper and lower nodes correspond to  the more general states for magnon excitations for the \(SU(2)_L\) wing (\(u\)-roots) and the \(SU(2)_R\) wing (\(v\)-roots).
}
 }
\end{figure}
%%%%%%%%%%%%%%%%%%%%%%
\Eq{Y:eq} has however many solutions and only one of them
really leads to the ground state energy. It is commonly believed that
certain other solutions there describe the excited states \cite{Dorey:1996re,Bazhanov:1996aq}. The
energy of the \(N\)-particle excited states is again given in terms of $Y_0$ but is
modified
\begin{equation}\label{E:eq}
E(L)=-\frac{1}{2}\int d\theta \cosh(\pi\theta)\log(1+ Y_0)+\sum_{j=1}^N m\cosh(\pi
\theta_j)\;,
\end{equation}
where the extra terms are inspired by the analytic continuation in
$L$ and the points $\theta_j$ \cite{Dorey:1996re} are singularities of the
integrand in the first term
\begin{equation}\label{Y0m1:eq}
Y_0(\theta_j\pm i/2)=-1\;,\qquad j=1,2,\ldots,N\;.
\end{equation}
 As we shell see, the last equation is nothing but the Bethe ansatz equation for physical
rapidities modified at the finite volume. The last term in \eqref{E:eq} is generated from the integral \eqref{E0:eq} by picking up the logarithmic poles \eqref{Y0m1:eq}.

Our goal in this section is to make use of the integrability of the Y-system
rewriting it in the form of classical integrable discrete Hirota dynamics. This
allows us to write down explicitly a solution for all $Y_n$ in terms of
a finite number of functions. Then we will restrict ourself to a
certain sub-class of physically relevant solutions with  particular analytic
properties. The analyticity will allow us to fix the functions
completely and parameterize all the physical solutions for the excited \(N\) particle states in terms of a
finite set of complex parameters, Bethe roots, restricted by supplementary Bethe equations reducing in the infinite volume to the usual Bethe equation.

\subsection{Hirota equations}

The $Y$-system equations \eq{Y:eq} can be seen as a gauge invariant
version of the so called Hirota equation or $T$-system\begin{equation}\la{eq:Hir}
T_k(x+i/2)T_k(x-i/2)-T_{k-1}(x)T_{k+1}(x)
=\Phi\left(x+i\frac{k}{2}\right)\bPhi\left(x-i\frac{k}{2}\right)\;.
\end{equation}
It can be easily checked  \cite{Kuniba:1993cn}
that Hirota equation is equivalent  to the
$Y$-system \eq{Y:eq} if we denote
\begin{equation}\label{TY}
Y_k(x)=\frac{T_{k+1}(x)T_{k-1}(x)}{\Phi\left(x+i\frac{k}{2}\right)
\bPhi\left(x-i\frac{k}{2}\right)}
.\end{equation}
 At first sight, this is just another trivial rewriting of
the TBA equations, however the Hirota form appears to be particulary useful. Using
Hirota equation we can also write
\beq\label{eq:1pY}
1+Y_k(x)=\frac{T_{k}(x+i/2)T_{k}(x-i/2)}{\Phi\left(x+i\frac{k}{2}\right)
\bPhi\left(x-i\frac{k}{2}\right)}\;.
\eeq
Let us point out here an important fact. By evaluating the above
equation for $k=0$ at $\theta_j\pm i/2$ where $\theta_j$ is a zero of
$T_0$ we observe that
\begin{equation}
T_0(\theta_j)=0\,\,\, \Rightarrow \,\,\, Y_0(\theta_j\pm i/2)=-1\,
\la{mif}
\end{equation}
which is the Bethe ansatz  \eq{Y0m1:eq}. We will use this fact
to associate zeroes of $T_0$ with physical rapidities.

Since $Y_k(x)$ are real functions by their physical meaning (for ground
state they are ratios of densities of complexes and of their holes, see
Appendix \ref{AppA}) we can restrict ourself to the case when $T_k$ are
real functions and $\Phi$ and $\bar \Phi$ are complex conjugated
functions.

Hirota equation (\ref{eq:Hir}) is integrable and has a Lax
representation through the auxiliary problem \cite{Krichever:1996qd}
\begin{eqnarray}\nonumber
T_{k+1}(x)Q\left(x+i\frac{k}{2}\right)-T_{k}\left(x-\frac{i}{2}\right)
Q\left(x+i\frac{k}{2}+i\right) &=&
 +\Phi\left(x+i\frac{k}{2}\right)\bar Q\left(x-i\frac{k}{2}-i\right)\\
T_{k-1}\left(x\right)\bar Q\left(x-i\frac{k}{2}-i\right)
-T_{k}\left(x-\frac{i}{2}\right)\bar Q\left(x-i\frac{k}{2}\right)
&= & -\bPhi\left(x-i\frac{k}{2}\right)
Q\left(x+i\frac{k}{2}\right)\;.
\label{LAX}
\end{eqnarray}
The compatibility of these two equations for the bi-vector of  functions $\{Q(x),\bar Q(x)\}$  leads to the initial Hirota equation. Here $\bQ$ is the complex conjugate function to $Q$.  Note that if $T_k(x)$ are real functions then the
second equation is simply the complex conjugate of the first one after
shifting $k\to k+1$ and $x\to x+i/2$. Two particularly
useful relations from this Lax representation are
\begin{eqnarray}\label{JUMPS}
T_{1}(x)&=&T_{0}(x-i/2)\frac{Q(x+i)}{Q(x)}+\Phi(x)
\frac{\bQ(x-i)}{Q(x)}\nn\,,\\
T_{-1}(x)&=&T_{0}(x+i/2)\frac{Q(x)}{Q(x+i)}-\Phi(x)\frac{\bQ(x)}{Q(x+i)}\;,
\end{eqnarray}
Note that the first relation in \eqref{JUMPS} is a generalization of the famous Baxter equation usually
written for the spin chains. We will see that in the infinite volume
limit $\Phi(x)=T_0(x+i/2)$ and that these equations  reduce to the usual
Baxter equation for spin chains, where $T_1$ plays the role of the
transfer matrix in fundamental representation for the magnons of the
$SU_R(2)$ wing of the theory, whereas the second equation plays a
similar role for the $SU_L(2)$ wing (see Fig.\ref{pic3}).

The main advantage of the Lax equations (\ref{LAX}) is that they are linear in
$T_k$ and we can easily express any $T_k$ in terms of $T_0,\Phi$ and $Q$ in the explicit form
\cite{Krichever:1996qd}\beqa\la{HSOL}
T_k(x)&=&\frac{Q\left(x+i\frac{k+1}{2}\right)}{Q\left(x-i\frac{k-1}{2}\right)}
T_0(x-ik/2)\\ \nn
&+&Q\left(x+i\frac{k+1}{2}\right)\bQ\left(x-i\frac{k+1}{2}\right)
\sum_{j=1}^k\frac{\Phi\left(x-i
\frac{k+1}{2}+ i j\right)}{Q\left(x-i\frac{k-1}{2}+i j\right)Q\left(x-i\frac{k+1}{2}+i j\right)}\;.
\eeqa
This  leads to a  quite general and explicit solution of the $Y$-system via \eq{TY}. A nice feature of
this form is that one can efficiently analyze the $L\to\infty$ limit  and
reproduce the asymptotic spectrum described by BAE
\eqs{BAEmiddle0}{BAEuv0}. This will be the goal of the next section.

Hirota and Lax equations exhibit several important symmetries. First of
all a discrete symmetry exchanging the $u$-wing and the $v$-wing (right and left \(SU(2)\)):
$Y_k \leftrightarrow Y_{-k}$ is induced by
\beq\label{wexch} T_k\leftrightarrow T_{-k}\;\;,\;\;
\Phi\leftrightarrow -\bar \Phi\;\;,\;\;
Q \leftrightarrow \bar Q^{--}\;\;,\;\;\bar Q \leftrightarrow Q^{++}\;,
\end{equation}
which will be quite useful for our further constructions. Moreover, both
equations (\ref{eq:Hir}) and (\ref{LAX}) are invariant under the gauge
transformation
\begin{eqnarray}\la{eq:GAUGE}
T_k(x)&\rightarrow& g\left(x+i\frac{k}{2}\right)\bar
g\left(x-i\frac{k}{2}\right) \,
T_k(x), \nn\\
  \Phi(x)&\rightarrow& g(x-i/2)g(x+i/2)\Phi(x), \,\nonumber\\
  \bPhi(x)&\rightarrow& \bar g(x-i/2)\bar g(x+i/2)\bPhi(x), \,\nonumber\\
  Q(x)&\rightarrow&g(x-i/2)Q(x).
%  \bar Q(x)&\rightarrow&\bar g(x+i/2)\bar Q(x)
\end{eqnarray}

To preserve the reality of $T_k$ we should assume that $\bar g$ is the
complex conjugated function to $g$. These transformations leave
 $Y_k(x) $ invariant.

The general solution of Hirota equation (\ref{eq:Hir}) can be also
presented in a determinant form \cite{Krichever:1996qd}
\begin{eqnarray}
T^{}_k(x)=h(x+ik/2)\left|
  \begin{array}{cc}
    Q(x+i\frac{k+1}{2}) & R(x+i\frac{k+1}{2}) \\
    \bar Q(x-i\frac{k+1}{2}) & \bar R(x-i\frac{k+1}{2})\\
  \end{array}
\right|\label{TQRG}
\end{eqnarray}
where $h(x)$ is a periodic function: $h^{++}\equiv h(x+i)=h(x)$ and
$Q,R$  are    two linearly independent solutions of the Lax equations
\eqref{LAX} related by the Wronskian relation
\begin{eqnarray}
\Phi_{}(x)=h(x+i/2)\left|
  \begin{array}{cc}
    R(x) & Q(x) \\
    R(x+i) & Q(x+i)\\
  \end{array}
\right|\label{PhiQRG}
.\end{eqnarray}This determinant form will be very useful when we
will formulate the general solution of the finite size PCF system for
any state. It is not absolutely necessary to use it, but it simplifies
some derivations.

\subsection{Asymptotic Bethe Ansatz and Classification of the Solutions\la{ABA}}\label{ASS}

The main problem in computing the exact spectrum of the $SU(2)$ PCF is
to find  the physical solutions to the $Y$-system (\ref{Y:eq}) or,
alternatively, to the Hirota equation (\ref{eq:Hir}), i.e., obeying the
right asymptotic properties (\ref{asympt}).
% and some certain analytic properties
%which we will discuss later
Their classification is a complicated task, especially when we want to
take into account not only the excitations of $U(1)$ sector but also
the ``magnon" type excitations of $SU(2)_L$ and $SU(2)_R$ sectors. The
goal of this section is thus to identify the large $L$ solutions to the
$Y$-system (\ref{Y:eq}). The discussion in this section is not
completely rigorous since our only goal is to get an idea of how asymptotic Bethe ansatz (ABA)
\eqs{BAEmiddle0}{BAEuv0} appears from the $Y$-system. Together with the expression
(\ref{energy}) the ABA equations must appear from the large $L$
asymptotic of exact solutions, as yielding the leading order value of
the full spectrum.

The main simplification in the large $L$ limit is  that $Y_0\to 0$.
From \eq{asympt} we see that $Y_0\to 2e^{- L\cosh(\pi
x)}$ and we are left with two decoupled chains of equations for $k>0$
and $k<0$ \cite{Balog:2003xd}. For each wing we can introduce two sets of \(T_k\) describing the corresponding solutions of the whole \(T\)-system:
$T_{k}^{u}$ and $T^v_k$ such that $Y_{k>0}$ ($Y_{k<0}$ )can be expressed in terms
of $T^u_k$ ($T^v_k$) by the formula \eqref{TY}. Then $Y_0=0$ implies
\beq
T^u_{-1}=0\;\;,\;\;T_{1}^v=0\;.
\eeq
Let us focus on $T^u_k$ since we can always use the wing exchange
symmetry (\ref{wexch}) to describe $T_k^v$ .

We have to identify the solutions which will lead  to the
asymptotic spectrum described by the ABA. It turns out that in terms of
Hirota potentials $T_k$ it is enough in this limit to make very simple assumptions, namely:
\begin{itemize}
\item All $T_{k>0}^u(x)$ are polynomials at \(L\rightarrow\infty\). We denote in this limit  $T_0^u(x)\approx\prod_j(x-\theta_j)     \equiv\phi(x).$
\item $Q^u(x)$ is a  polynomial with real roots which we denote $Q^u(x)=\prod_j(x-u_j)$.
\end{itemize}
Then from \eq{JUMPS} we see that
\beq\la{LinfT1}
\Phi^u(x)=T^u_{0}(x+i/2)\;\;\text{and}\;\;T^u_1(x)=\frac{T_0^u(x+i/2)Q^u(x-i)+T_0^u(x-i/2)Q^u(x+i)}{Q^u(x)}\;.
\eeq
From the polynomiality condition for $T_k^u(x)$   and \(T_k^v(x)\) we  get
precisely the auxiliary Bethe equations \eq{BAEuv0}.

Finally, we should note that eq.(\ref{BAEmiddle0}) for the physical
rapidities $\theta_j$ is also satisfied. This  follows from imposing
$Y_0(\theta_j\pm i/2)=-1$ for all zeros $\theta_j$ of $T_0^u$, see
(\ref{mif}). At first sight, this seems to  be impossible to satisfy since, as we noticed,
$Y_0(x)$ is small. However this smallness appears because $Y_0$ is proportional to $e^{-L\cosh(\pi
x)}$ which is indeed small inside the physical strip  $-1/2<\im x<1/2$ but is of order $1$ on the boundary of this strip.
To impose this condition we must first compute $Y_0$ to the next order.

%Obviously, we can repeat the same arguments for the function $T_{-1}^v$.

From (\ref{Y:eq}) at $n=0$ we get
\begin{equation}
Y_0^+ Y_0^-=
\frac{T_1^{u+}T_{-1}^{v+}T_1^{u-}T_{-1}^{v-}}{(\phi^{++}\phi^{--})^2}\;.
\end{equation}
Defining $S(x)=\prod_{j=1}^{N} S_0(x-\theta_j)$ we have, from the
crossing relation (\ref{crossing}), $ S^{++} S = {\phi}/{\phi^{++}} $ ,
so that
\begin{equation}
Y_0^+ Y_0^-= \left(\frac{T_1^{u}(x)T_{-1}^{v}(x)
S^2(x+i/2)}{\phi^2(x-i/2)}\right)^+\left(\frac{T_1^{u}(x)T_{-1}^{v}(x)
S^2(x+i/2)}{\phi^2(x-i/2)}\right)^-\;,
\end{equation}
from which we can identify $Y_0$ up to a zero mode factor of $y_0=e^{-
L \cosh \pi x}$ which obeys $y_0^+y_0^-=1$. Such factor should be
included into $Y_0$ to ensure the proper asymptotic (\ref{asympt}).
Thus we find
\begin{equation}\label{LargeLY}
Y_0(x)\simeq e^{-L\cosh(\pi x)}T_1^u(x)
T_{-1}^v(x)\frac{S^2(x+i/2)}{\phi^2(x-i/2)}\,.
\end{equation}
Evaluating it at $x=\theta_k-i/2$ and using \eq{LinfT1} we get
\begin{equation}
-1\simeq e^{i L\sinh(\pi \theta_k)}\frac{\Q_u(\theta_k+i/2)\Q_v(\theta_k+i/2)}{\Q_u(\theta_k-i/2)\Q_v(\theta_k-i/2)}\prod_j S_0^2(\theta_k-\theta_j)\,,
\end{equation}
which is nothing but  the main ABA equation
(\ref{BAEmiddle0}) for the middle node in fig.\ref{pic3}.  We use here the notations
\beq\label{eq:QcQ}
\Q_v(x)=\bar Q_v(x-i)\;\;,\;\;\Q_u(x)=Q_u(x)\;,
\eeq
to make the \(u\)- and \(v\)-wings more symmetric. The advantage of these notations is that the wing exchange symmetry \eq{wexch}
simply exchanges $\Q_v$ and $\Q_u$ and in the large $L$ limit they are
real polynomials.

Finally, since $Y_0(x)$ is exponentially suppressed for real $x$ we can
drop the integral contribution in (\ref{E:eq}) which leaves us with the
energy as a sum of  energies of individual particles, precisely as
expected from (\ref{energy}).

Notice that the Zamolodchikov asymptotic scattering theory is implicitly contained in the \(Y\)-system, as we see from the appearance of the scalar scattering factor \(S^2\) in the formula
\eqref{LargeLY}.

%%%%%%%%%%%%%%%%%%%%%%%%%%%%%%%%%%%%%%
\subsection{Probing the finite volume}

Now, having established the solution at infinite volume, we need an
insight into the analytic properties of \(T\)-functions  in a finite,
though large, volume. Let us find perturbatively the finite $L$
corrections for the simplest vacuum solution which for large $L$
corresponds to $Q^u=Q^v=1,\,\phi=1$. From \eq{HSOL} one can see that
for this case, to the leading order, $T^u_k\simeq k+1$  which implies for
$Y_k\simeq |k|^2+2|k|$. Thus we are looking for a solution in the form
\begin{equation}
Y_k=|k|^2+2|k|+y_k\;\;,\;\;k=-\infty,\dots,\infty
\end{equation}
where the first two terms in the r.h.s. are the trivial solution at
$L=\infty$, where as  $y_k\sim Y_0$ are small. We will see that the
solution for the perturbation is unique under the assumption that when
$k\to\infty$ the perturbation goes to zero $y_k\to 0$.
  The linearized $Y$-system in the Fourier form is
\begin{equation}
\frac{ k}{k+2}\tilde s\,\tilde y_{k+1}-\tilde y_k+\frac{k+2}{k}\tilde s\,\tilde y_{k-1}
=0\;\;,\;\;k\geq0\;
\end{equation}
where $\tilde y_k$ is the Fourier transform of $y_k$ and  $\tilde
s=\frac{1}{2\cosh(\omega/2)}$ is the Fourier transform of the kernel
$s=\frac{1}{2\cosh(\pi\theta)}$. $\tilde y_0=\tilde Y_0$ is a fixed
function. We see that this is a second order recurrence equation which
in general has two linear independent solutions. Fortunately it can be
solved explicitly.\footnote{One can use
\textrm{RSolve} function in
\textsl{Mathematica} to find the solution.} The general solution reads
\begin{equation}\nn
\tilde y_k=\frac{k(k+1)(k+2)}{2}\left(\[\frac{e^{-\frac{k|\omega|}{2}}}{k}- \frac{e^{-\frac{(k+2)|\omega|}{2}}}{k+2}\]C_1(\omega)+
\[\frac{e^{\frac{k|\omega|}{2}}}{k}- \frac{e^{\frac{(k+2)|\omega|}{2}}}{k+2}\]C_2(\omega)\right)\, .
\end{equation}
 The needed solution satisfying $\tilde y_0=\tilde
Y_0\;,\;\tilde y_{\infty}=0$ corresponds to $C_1=\tilde Y_0,\;C_2=0$.
Making the inverse fourier transformation we get
\begin{equation}
y_k = \frac{k(k+1)(k+2)}{\pi}\left(\frac{1}{4x^2+k^2}
-\frac{1}{4x^2+(k+2)^2}\right)* Y_0\;.
\end{equation}
It can be easily checked that the approximate $T_k$ yielding this solution through (\ref{TY}) are
\begin{equation}\la{TlargeY}
T^{u}_{k-1}=T^{v}_{1-k}\simeq k+\frac{k/\pi}{4x^2+k^2}*
Y_0\;\;,\;\;k\geq0\;.
\end{equation}
and
\begin{equation}
\Phi(x)=1+\frac{1/\pi}{4(x+i/2+i0)^2+1}*Y_0 \,.
\end{equation}
The $i0$ in this expression can be dropped when computing $Y_{k>0}$ from  (\ref{TY}) but is included in this expression so that (\ref{TY}) can also be used for $k=0$, for more details see the discussion in the next subsection.

An important feature of this asymptotic solutions for $T_{k}$, which
should persist at any $L$, is that it acquires two branch cuts at $x\in
{\bf R}\pm i k/2$ when  $L\to\infty$.\footnote{The term ``branch cut"
is not very appropriate here since the infinite cut has no branch
points. However, as we shall see, a spectral representation will allow
us to define $T_{k}(x)$ in the whole complex plane in terms of
spectral density integrals along the cuts.}

\subsection{Exact solution for the vacuum}

We will now extend the solution found in the previous section to
arbitrary $L$. First, we notice that the solution in terms of $T_k$ is
much simpler than in terms of $Y_k$. For the vacuum   we can
use the following ansatz inspired by \eq{TlargeY}
\begin{equation}\la{YT1}
T_{k-1}= k+\frac{k/\pi}{4x^2+k^2}* f,\qquad k=+0,1,2,\dots
\end{equation}
where $f$ is some function which for large $L$ becomes $Y_0$.
 One can easily see from the linear system \eq{LAX}
at $Q=\bar Q=1$ that this ansatz solves the Hirota equation and can be presented in the form
\eq{HSOL} with \(\Phi(x)=T_0(x+i/2+i0)\). Thus the Y-system equations
\eq{Y:eq} for $|k|\geq 2$ are satisfied automatically. Notice that
none of the $T_k$'s has singularities on the real axis, which is of
course a necessary feature of the solution: the physical quantities $Y_k$ should not be singular there.

To check that the equation for $k=1$ is also satisfied we have to define $Y_0$ in terms
of $T_k$. For that we can simply analytically continue
\eq{YT1} to the point $k=+0$ which gives $T_{-1}(x)=f(x)/2$. We also have
 $\Phi(x)=T_0(x+i/2+i0)$, $\bPhi(x)=T_0(x-i/2-i0)$ as mentioned above. These properties are supported by the second  equation \eqref{JUMPS} which can be viewed as yielding the spectral density in terms of a jump on any of two infinite cuts. Then we get
\begin{equation}\la{eqY0}
Y_0(x)=\frac{T_{0}(x+i/2-i0)T_{0}(x-i/2+i0)}{T_{0}(x+i/2+i0)T_{0}(x-i/2-i0)}-1=\frac{T_{1}(x)f(x)/2}{T_{0}(x+i/2+i0)T_{0}(x-i/2-i0)}
\end{equation}
This equation relates $Y_0$ and $f$. With $Y_0$ so defined the Y-system
equations at $|k|=1$ are now also satisfied. However the equation (\ref{Y:eq0}) for $k=0$
is still not used. Using 
$$(1+Y_1)(1+Y_{-1})=(1+Y_1)^2=\left(\frac{T_1(x+i/2)T_1(x-i/2)}{T_0(x+i)T_0(x-i)}\right)^2$$
and recalling that $s$ is the inverse shift operator we obtain\footnote{We introduce a natural notation
$g^{*s}\equiv e^{s*\log g}$.}
\begin{equation}\la{eqY01}
Y_0(x)=e^{-L\cosh(\pi x)}\frac{T_1^2(x)}{
\[T_0(x+i)T_0(x-i)\]^{*2s}}.
\end{equation}
Combining it with \eq{eqY0} we  get
\beq\la{VacIt}
f(x)=2T_1(x)\frac{T_{0}(x+i/2+i0)T_{0}(x-i/2-i0)}{
\[T_0(x+i)T_0(x-i)\]^{*2s}}e^{-L\cosh(\pi x)}\;,
\eeq
which,  in virtue of  the \eq{YT1}, gives a closed equation for $f(x)$.

Notice that from \eq{VacIt} $T_{-1}(x)=f(x)/2$ is exponentially small
for large $L$ with
\begin{equation}\la{eqASSTm1}
T_{-1}(x) \simeq 2e^{-L\cosh(\pi x)}.
\end{equation}
The finite $L$ solution to equation (\ref{VacIt}) can be easily found by iterations, starting from this   large $L$ asymptotic and gradually diminishing \(L\).
We solved this equation numerically and get a perfect match with the existing results (see the Tab.\ref{tb1} comparing our results with \cite{Balog:2003yr}).

%%%%%%%%%%%%%%%%%%%
%%%%%%%%%%%%%%%%%%%
\begin{table}[h!b!p!]
$$
\bea{llll}
L& \text{Leading order}  & \text{\Eq{VacIt}} & \text{Results of
\cite{Balog:2003yr} }\\ \hline
L=4&-0.015513  & -0.015625736& -0.01562574(1)\\
L=2&-0.153121  & -0.162028968 & -0.16202897(1)\\
L=1  &-0.555502  & -0.64377457 & -0.6437746(1)\\
L=1/2&  -1.364756  & -1.74046938&  -1.7404694(2)\\
L=1/10& -7.494391  & -11.2733646&  -11.273364(1)
\eea
$$
\caption{\label{tb1}\it We solve numerically \eq{VacIt} the use $Y_0$ from
\eq{eqY0} to compute the energy of the ground state using
\eq{E0:eq}.}
\end{table}
%%%%%%%%%%%%%%%%%%%
%%%%%%%%%%%%%%%%%%%

In the next subsection, we generalize this solution to the
excited states in the \(U(1)\) sector.

\subsection{Generalization to $U(1)$ sector}\label{subsec:U1}
 In this section we will study in detail the $U(1)$ sector of the theory where we consider the states with $N$ particles with the same polarization, i.e. with no magnon excitations. Hence we can put all  $Q=1$. As mentioned before -- see \eq{mif} -- for $N$ particle states we expect
$T_0(\theta_j)=0$ for each of $N$ rapidities of the particles $\theta_{1},\dots,\theta_N$.

In the previous section the vacuum state, with no particles excited, was analyzed. We saw that $T_0(x)$ inside the physical strip, $\Phi(x)$ above the strip and $\bar\Phi(x)$ below the strip could be described by a single function
\beq
F_{}(x)=1+\frac{1/\pi}{4x^2+1}*T_{-1} \,, \la{Fvacuum}
\eeq
such that
\beq
F(x)=\left\{
\begin{array}{ll}
\Phi(x-i/2) &, \im(x)>1/2 \\
T_0(x) &, |\im(x)|<1/2 \\
\bar\Phi(x+i/2) &, \im(x)<-1/2
\end{array}
\right. \;.\la{Fdecomp}
\eeq
\begin{figure}[ht]
\epsfxsize=8cm
\centerline{\epsfbox{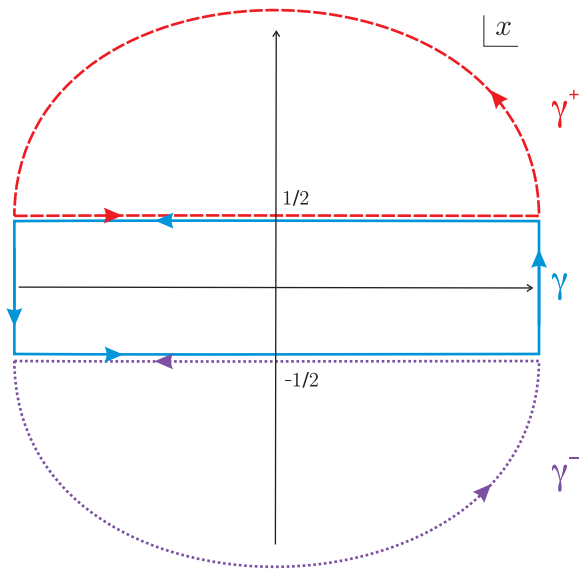}}
\caption{\label{FS}
\textsl{The function $F(x)$ in (\ref{ansatz}) can be recast as a contour integral as in (\ref{ansatz2}) with the contours as represented in this figure.
} }
\end{figure}
Here we build a generalization of (\ref{Fvacuum}) for the case when
$T_0$ has an arbitrary number of zeroes inside the physical strip
for which (\ref{Fdecomp}) holds:
\begin{equation}
F(x)=\phi(x)\left(1-
\int_{-\infty}^{\infty} \left(\frac{1}{\phi(y -i/2)}\frac{1}{x-y+i/2}
-\frac{1}{\phi(y+i/2)}\frac{1}{x-y-i/2}\right)\frac{T_{-1}(y)dy}{2\pi i} \right)\,, \la{ansatz}
\end{equation}
with  $\phi(x)\equiv \prod_{j=1}^N(x-\theta_j)$.
The overall factor of $\phi(x)$ appears because $T_0(\theta_j)=0$.
The spectral representation of $F(x)$ as two integrals over the two infinite cuts at $\im(x)=\pm 1/2$ is inspired by (\ref{Fvacuum}) and can be also seen from the linear problem (\ref{LAX}). Indeed, we have
\beq
T_{-1}(x)=T_0(x+i/2)-\Phi(x)=T_0(x-i/2)-\bar \Phi(x)
\eeq
which justifies the choice of spectral densities used in (\ref{ansatz}). To see that (\ref{Fdecomp}) indeed holds we write (\ref{ansatz}) as
\begin{equation}
F(x)=\phi(x) \left(
\ccw\oint_\gamma \frac{dy}{2\pi i}\frac{T_0(y)/\phi(y)}{y-x}+
\ccw\oint_{\gamma^+} \frac{dy}{2\pi i}\frac{\Phi(y-i/2)/\phi(y)}{y-x}+\ccw\oint_{\gamma^-} \frac{dy}{2\pi i}\frac{\bar \Phi(y+i/2)/\phi(y)}{y-x}\right)\,. \la{ansatz2}
\end{equation}
The contours $\gamma$, $\gamma^+$ and $\gamma^-$ encircle respectively the physical strip, the region above the strip and the region below the strip, see figure \ref{FS}. For this relation to be equivalent to (\ref{ansatz}) we require that for large $x$ we should have $T_0(x),\Phi(x-i/2),\bar \Phi(x+i/2) \to \phi(x)$ at \(|x|\to \infty\) along the corresponding contour. Finally, for (\ref{Fdecomp}) to hold, the ratios in (\ref{ansatz2}) should be analytic inside the corresponding contours. Notice that at large $L$ the function $T_{-1}$ is exponentially small and thus $\Phi^-,T_0,\bar\Phi^+ \to  \phi(x)$ as expected from our discussion in section \ref{ASS}, to get the ABA equations. The large $x$ limit
should be similar to the large $L$ limit since the source term in the $Y$-system
$e^{-L\cosh(\pi x)}$ is small in both cases.

Let us now consider the other Hirota functions $T_k$. From (\ref{LAX}) we have
$T_1(x)=T_0(x+i/2)+\bar\Phi(x)=\Phi(x)+T_0(x-i/2)$ which in terms of the function $F(x)$ reads
\beq
T_1(x)=F(x+i/2+i0)+F(x-i/2+i0)=F(x+i/2-i0)+F(x-i/2-i0)\;,
\eeq
so it is indeed regular on the real axis. Notice that $T_1$ is regular at
least inside the enlarged strip $|\im(x)|<1$. In the same way we can easily see that $T_{k>0}$ is analytic inside the strip $|\im(x)|<\frac{k+1}{2}$.

Having expressed $T_0$, $\Phi$ and $T_1$ in terms of $T_{-1}$ through the function $F(x)$ we can find a closed equation of $T_{-1}$ from the $Y$-system equation for $n=0$. The derivation is parallel to  the one in the previous section and it leads to\beq
T_{-1}(x)=(F(x+i/2)+F(x-i/2)) \frac{F(x+i/2+i0)F(x-i/2-i0)}{
\[F(x+i)F(x-i)\]^{*2s}}\,e^{-L\cosh(\pi x)}\;, \la{eqF}
\eeq
supplemented by the quantization condition $Y_0(\theta_j+i/2)=-1$. As before, the solution to these equations can be easily found from iterations as is
explained in the Sec.\ref{sec:NUM}. The  numerically calculated energies of a few states of this \(U(1)\) sector are presented on the fig.\ref{pic4}.

In the next section, we generalize these results to any
excited states including the magnon polarizations. We will use a different strategy and incorporate the gauge
invariance of $Y$-system to find the solutions
 of Y-system \eq{Y:eq} matching the $L=\infty$
asymptotic of the Sec.\ref{ABA}.

\section{Finite size  spectrum for a general state of PCF }\label{GEN}

We will now describe how to construct the solution for the most
general state of the PCF at finite volume $L$, having an arbitrary number of
physical particles with arbitrary polarizations in the $SU(2)_R$ and
$SU(2)_L$ wings (characterized by left and right ``magnons" $u_i$ and $v_i$). Our
method is based on the following observations and steps:

\begin{itemize}
\item  {We know from \eq{LargeLY} the structure of the poles and zeroes of all $Y_k$'s in
the limit $L\to\infty$ when $Y_0=0$. We assume that this
structure will qualitatively persist even for finite $L$, and
the classification of the appropriate solutions of the Y-system
will follow the same
 pattern of poles and zeroes.}

\item {We will recast the Y-system in terms of T-system (Hirota equation) since the
analytic structure of  $T_{k}$'s is much simpler than of $Y_s$
as we saw from the vacuum solution
 \eqref{TlargeY} at  $L\to\infty$.}

\item{For any ``good" solution of $Y$-system there is a family of solutions of
  Hirota equations related by  gauge transformations
  \eqref{eq:GAUGE}. Hirota equation can be solved explicitly in terms of
$T_0,\Phi$ and $Q$ as in \eq{HSOL}.}

\item{For $L\rightarrow\infty$ we have two independent solutions for $T_k$'s as we saw in the previous section. For one solution $T^u_k$ are asymptotically polynomials for $k> 0 $ and for another one $T^v_k$ with $k< 0$ are polynomials when $L$ is large. We can then smoothly continue these two solutions to finite $L$'s using the gauge freedom to preserve polynomiality of $Q$'s.}

 \item{ We have two global solutions of
Hirota equation which can be  parameterized by $T_0^u,\Phi_u,Q^u$
, and by $T_0^v,\Phi_v,Q^v$.      They represent however
the same and unique solution of the Y-system and thus should be related by a gauge
transformation $g:\;T^v_s=g\circ T_{s}^{u}$, see \eqref{eq:GAUGE}.}

\item{   Using certain assumptions about analyticity of $T_0^{u}$ and $\Phi_{u}$, supported by the Lax equations \eqref{LAX}, we can express them as different analytic branches of the same analytic function $G_{u} $. The same can be done for  $T_0^{v}$ and $-\bar\Phi_{v}$ in terms of $G_v$.
 }

%\item {We can always choose a gauge for each of these parametrizations
%in such a way as to make $Q^u$ or $Q^v$  polynomial at any $L$, as in
% \eqref{PolyQF} Their roots does not have to be real  at a finite size L. % In the limit
% $L\to\infty$ this polynomials will smoothly transform into the
% polynomials $Q^u$ or $Q^v$ , depending on our choice to
% describe the right or the left magnons. The roots of  $\phi,Q^u,Q^v$ in %\eqref{PolyQF}    will be fixed by  the finite size analogues of Bethe ansatz % equations.}

\item {The solution will be completely fixed by the  existence
of such a gauge transformation $g(x)$ which relates its $u$-- and the $v$--representations. At the end we will have
one
 single non-linear integral equation (NLIE) on $g(x)$.}

\end{itemize}

The final equation for  $g(x)$ is new for the Principal Chiral Field. It is different  from the system of 3 DdV type equations used for the same model in \cite{Hegedus:2004xd}. Still it resembles in many aspects the non-linear Destri-de~Vega (DdV) equation which appears when studying other integrable models. Indeed, our method is very general and it allows to generate DdV-like equations for large classes of sigma models in a systematic way. For the models for which a DdV equation is known we expect our integral equation to coincide with it after an appropriate change of variables. We check this hypothesis on the $SU(2)$ chiral Gross-Neveu model for which we re-derive indeed the known integral equation.

\subsection{Exact  equations for the finite volume spectrum }

In this section, we will derive the finite volume   spectral  equations of the previous section in the most general form, valid for all excited states of the model with any number of physical particles with arbitrary polarizations (i.e. with any quantum numbers).

As we discussed below in the infinite volume, the solution of Y-system with
$Y_0=0$ can be described in terms of two independent sets of Hirota potentials $T_k^u$ and $T_k^v$. Since these two different solutions of Hirota equation
correspond to the same solution of Y-system they are related by a gauge transformation
$g(x)$. These  two solutions of Hirota equation can be continuously and unambiguously deformed all the way from very large \(L \), where we know the solution (see the previous section), to  any finite
value of $L$. The gauge ambiguity for any of the two solutions,  $T_k^u$ or $T_k^v$, can be fixed by choosing
$Q_u$ and $Q_v$ to be polynomials for any $L$. Of course we can no longer
assume $T_k^u$ and $T_k^v$, as well as the corresponding $\Phi_v$ and $\Phi_u$, to be polynomials. Instead we will assume certain analytic properties for them and we will see their consistency with the
solution we find at the end.

We
introduce a polynomial $\phi(x)$ with real zeroes
$\theta_j,\,j=1,2,\ldots,N$  of $T^u_0$. They correspond to the rapidities of physical particles on the circle. The gauge function \(g(x)\) relating the two solutions of the \(T\)-system is assumed to be  regular  and to have no zeros on the physical strip, so that
$T_0^v=g\bar gT_0^u$ has the same zeroes as $T^u_0$  there. We also assume that\footnote{When \(x\rightarrow\infty\) we know that  \(Y_{0}(x)\rightarrow e^{-L\cosh(\pi x)}  \) , i.e. it is exponentially small, as in the case of large \(L\), and the Y-system, as well as the T-system splits in to two independent \(u\)- and \(v\)-wings with \(T_{0}(x)\sim\Phi(x)\sim\phi(x)\sim x^{N}.\  \)  }
\begin{itemize}
\item{$\frac{\Phi_u(x)}{\phi(x+i/2)}$ ($\frac{\bPhi_u(x)}{\phi(x-i/2)}$)
is regular for $\im x>-1/2$ ($\im x<1/2$) in the whole upper
 (lower) half plane and goes to $1$ for $|x|\rightarrow\infty$
 in all directions in the upper (lower) half plane;}
\item{$\frac{T_0^u(x)}{\phi(x)}$ is regular and goes to $1$ at $x\rightarrow\pm\infty$ inside the physical strip $-\half<\im x< \half$; }
\end{itemize}

The first property is somewhat similar to the forth property from the previous subsection: the large \(x\) asymptotics is governed by the same exponential \(e^{-L\cosh\pi\th}\) as the large \(L\) asymptotics. As a consequence of the second assumption,  inspired by the integral representation \eqref{ansatz} for the \(U(1) \) sector, $T_{k>0}^u(x)$ are regular for $-(k+1)/2<\im x<(k+1)/2$ .

Similarly, for another solution we assume that
\begin{itemize}
\item{$\frac{\bPhi_v(x)}{\phi(x+i/2)}$ ($\frac{\Phi_v(x)}{\phi(x-i/2)}$) is regular
for $\im x>-1/2$ ($\im x<1/2$) in the all upper (lower) half
plane and goes to $1$ when \(|x|\to\infty \) in all directions in the upper (lower) half plane}
\item{$\frac{T_0^v(x)}{\phi(x)}$ is regular and goes to $1$ at $x\rightarrow\pm\infty$  inside
 $-\half<\im x<\half $ strip.}
%\item{$T_{-k<0}^v(x)$ are regular for $-(k+1)/2<\im x<(k+2)/1.$}
\end{itemize}

%These assumptions need some justification. For example, if \(|x|\rightarrow\pm\infty\) % then \(Y_{0}\rightarrow 0   \), \(T_{-1}^{u,v}\rightarrow0\) and the roots %\(u_j\) become real.   Then due to \eqref{JUMPS} \(\Phi^{u,v}\approx T_{0}^{u,v+} %\). This limit is similar to the \(L\rightarrow\infty\) limit, thus, with %the exponential precision \com{V: not sure...},  \(\Phi^{u,v}\approx T_{0}^{u,v+} %\rightarrow\phi(x)\approx x^{N} \).  This proves the first two conditions %for each wing. The last condition is natural  from the definition \eqref{eq:1pY} % of \(Y_{0} \)    since we want \(Y_{0} \) be nonsingular in the physical %strip  \(-1<\im x<1\).

Note that with these properties all $Y_{k\ne 0}\,$ are automatically
analytic in the physical strip, as we see from \eqref{TY}.
\(Y_0\) is, in the strict sense, only analytic on the real axis, but the detailed analysis of the Appendix \ref{ANALYT} shows that we can expect its analyticity even in a finite strip around the real axis. As concerns the \(T\)-functions, although we  use  in different circumstances $T_k^u$   or  $T_k^v$,
we will get the same result for all $Y_k$, since they are related by
a gauge transformation $g(x)$. But analyticity will be explicit only
for one wing at a time: at \(k\ge 1 \) for $T_k^u$   and at \(k\le 1 \) for $ T_k^v$.

The listed properties are enough to explicitly relate  $T_k^u$ , $T_k^v$ with the corresponding $\Phi_v$ and $\Phi_u$ using a certain integral representation for them. In Appendix \ref{GenSol2} we follow this line of arguments to formulate the complete set of equations for an arbitrary state, including a NLIE for \(g(x)\) and the associated  equations for the Bethe roots. However, it appears to be more advantageous, especially for the numerics, to use the integral representations for the logarithms of $T_k^u,T_k^v,\Phi_v,\Phi_u$. We will derive in what follows the corresponding equations defining the energy of a general state.

\begin{figure}[ht]
\epsfxsize=13cm
\centerline{\epsfbox{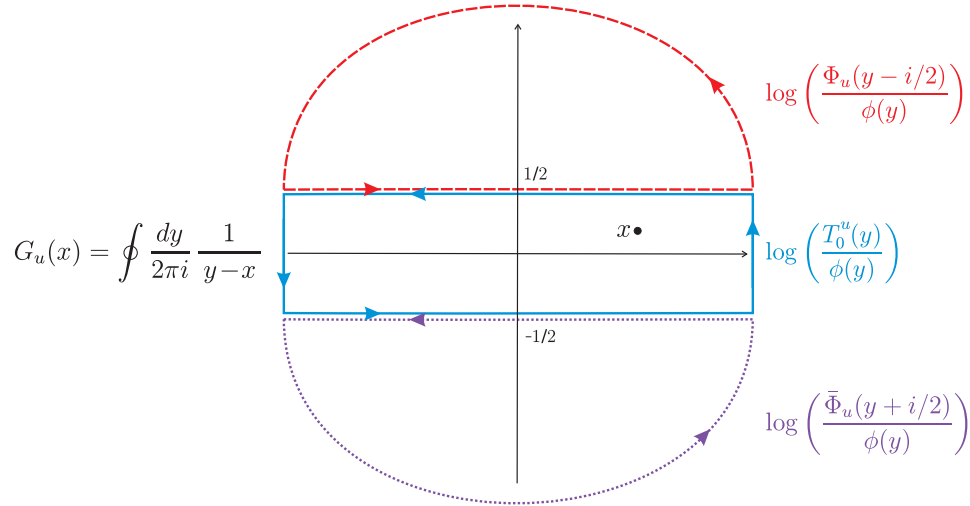}}
\caption{\label{contours}
\small\textsl{The ratios indicated close to each contour are analytic inside the corresponding contour. Thus we can obtain them in each of these regions using a single resolvent $G_u(x)$ as in (\ref{eq:RESOLV}).
} }
\end{figure}

Let us define  two new  functions analytic on a Riemann surface with two infinite cuts at $\im x=\pm i/2 $
\begin{equation}\label{eq:RESOLV}
G_w(x)=\frac{1}{2\pi i}\int_{-\infty}^{\infty}
\frac{\rho_w(y)}{x-y-i/2}dy-\frac{1}{2\pi i}\int_{-\infty}^{\infty}
\frac{\bar\rho_w(y)}{x-y+i/2}dy,\quad\;\;(w=u,v)
\end{equation}with the following spectral densities along the infinite cuts\footnote{These spectral densities denoted by $\rho$ should not be confused with the densities of Bethe roots $\varrho$ used in appendix \ref{AppA} in the derivation of the Y-system ground state equations.}
\beq\label{rhoUV}
\rho_u(x)=\log\left(\frac{T^u_0(x+i/2)}{\Phi_u(x)}\right),\quad\rho_v(x)=\log\left(\frac{T^v_0(x+i/2)}{-\bar\Phi_v(x)}\right),\eeq
and
their complex conjugates
\beq\label{rhoUVbar}
\bar\rho_u(x)=\log\left(\frac{T^u_0(x-i/2)}{\bar\Phi_u(x)}\right),\quad
\;\;
\bar\rho_v(x)=\log\left(\frac{T^v_0(x-i/2)}{-\Phi_v(x)}\right)\;.
\eeq
Then from (\ref{TY}) we have
\beq\la{Y0RHO}
\log(1+Y_0)=\rho_u+\bar \rho_u=\rho_v+\bar \rho_v\;.
\eeq
When $L$ is large enough we know from the results of the section \ref{ASS} that  $\frac{}{}T^w_0(x)\simeq \Phi_w(x-i/2)\simeq\bPhi_ w(x+i/2)\simeq\phi(x) $  and
thus to the leading order $\rho_{w}(x)$'s are exponentially small. It is also important for our analyticity assumptions listed above to hold that \(\rho_{w}(x)\sim e^{-L\cosh\pi x}\) at \(x\to\pm\infty\) along the real axis. Together
with these analyticity assumptions the following formulae
are true
at any \(L\)\begin{equation}\label{eq:GPhiT}
G_v(x)= \left\{
\bea{ll}
\log\displaystyle \frac{-\bPhi_v(x-i/2)}{\phi(x)}&\IM x>+1/2\\
\log\displaystyle \frac{ T^v_0(x)}{\phi(x)}&|\IM x|<1/2\\
\log\displaystyle \frac{-\Phi_v(x+i/2)}{\phi(x)}&\IM x<-1/2
\eea
\right.\;\;,\;\;
G_u(x)= \left\{
\bea{ll}
\log\displaystyle \frac{\Phi_u(x-i/2)}{\phi(x)}&\IM x>+1/2\\
\log\displaystyle \frac{T^u_0(x)}{\phi(x)}&|\IM x|<1/2\\
\log\displaystyle\frac{\bPhi_u(x+i/2)}{\phi(x)}&\IM x<-1/2
\eea
\right.\;.
\end{equation}

These formulas are easily understood from simple contour manipulation as depicted in figure   \ref{contours}. Let us consider the resolvent $G_u$, plug \eqref{rhoUV} and \eqref{rhoUVbar}
into \eqref{eq:RESOLV} and consider separately the terms containing $\frac{T_{0}^u(x)}{\phi(x)}$, $\frac{\Phi_u(x-i/2)}{\phi(x)}$ and $\frac{\bar\Phi_u(x+i/2)}{\phi(x)}$.  Since  $\frac{T_{0}(x)}{\phi(x)}\rightarrow 1,\,x\rightarrow \pm\infty$, we can close the contour in the integrals containing  $\frac{\log T_{0}(x)}{\phi(x)}$ around the physical strip and contracting it around the pole $y=x$ we obtain the middle relations in \eqref{eq:GPhiT} if $x$ lies in the physical strip. Similarly, using the fact that in the upper half-plane   $\frac{\Phi_u(x- i/2)}{\phi(x)}\rightarrow 1,\,x\rightarrow \infty$, we can close the contour in the integrals containing  $\log \frac{\Phi_u(x- i/2)}{\phi(x)}$ (after the obvious shift of integration variable) around the upper half-plane and contracting it around the pole $y=x $ we obtain the upper relations in \eqref{eq:GPhiT} provided ${\rm Im}(x)>1/2$. The integrals containing $\log \frac{\bar \Phi_u(x- i/2)}{\phi(x)}$ are treated similarly with the contours being closed in the lower half-plane.
For the resolvent $G_v$ the same sort of reasonings apply.

As we mentioned above, the two solutions of Hirota equation we defined in
this way, are related by a gauge transformation $T_k^v=g\circ T_{k}^u$. However,
the polynomials $Q_u$ and $Q_v$ are not necessarily related by this gauge
transformation $g(x)$. Instead, one can easily see that $Q_v$ is mapped to another linearly independent solution of
 \eq{LAX} $R^u=\frac{Q^v}{g^-}$. We can use \eqs{TQRG}{PhiQRG} to express all $T_k$'s and $\Phi$'s in
terms of \(Q\)'s.
In particular, we have
\begin{eqnarray}
\Phi_{u}&=&h^+\left(\frac{\Q_u^{++}\bar \Q_v^{--}}{g^-}-\frac{\Q_u \bar \Q_v}{g^+}\right)\;\;,\;\; T_0^{u}=h\left(\frac{\Q_u^{+}\Q_v^{+}}{\bar g}-\frac{\bar \Q_u^{-} \bar
\Q_v^{-}}{g} \right)\;.
\end{eqnarray}
%The other useful relations are
%\begin{eqnarray}\la{qBaxter}
%T_1^{u}&=&h^+\left(\frac{\Q_u^{++}\Q_v}{\bar g^-} -\frac{\bar \Q_u^{--} %\bar
%\Q}{g^+} \right)\;\;,\;\;
%T_{-1}^u=h^+\left(\frac{\Q_u\Q_v^{++}}{\bar g^+} - \frac{\bar \Q_u \bar %\Q_v^{--}}{g^-} \right).
%\eeqa
%Defining the quantities which will be central to our DdV-like equations
%\beq\la{eq:RHOu}
%A_\theta=-\frac{q_u^+\bar q_v^+}{\bar q_u^- q_v^-}\frac{g}{\bar g}
%\,\, ,\quad\ \,\, A_{h}^+=-\frac{q_u^{++}q_v^{--}}{q_u q_v}\frac{g^+}{g^-}
%\eeq
Similar relations for $v$ wing can be obtained from the gauge transformation
$\Phi_{v}=g^-g^+\Phi_{u}$ and $T_0^{v}
=\bar ggT_{0}^{u}$. For the densities (\ref{rhoUV}-\ref{rhoUVbar}) this yields\beq\la{eq:RHOu}
e^{\rho_u}=+\frac{T_0^{u+}}{\Phi_u}=\frac{\frac{g^{+}}{\bar g^+}\Q^{++}_u \Q_v^{++}- \bar \Q_u\bar\Q_v}{\frac{g^{+}}{g^-}\Q_u^{++}\bar \Q_v^{--}-\Q_u\bar\Q_v}
\;\;,\;\;
e^{\rho_v}=-\frac{T_0^{v+}}{\bar\Phi_{v}}=\frac{\frac{ g^+}{\bar g^+} \Q_v^{{++}}
   \Q_u^{{++}}-\bar\Q_v \bar \Q_u }{\frac{\bar  g^-}{\bar g^+}\Q_v^{{++}}\bar{\Q}_u^{{--}}
   - \Q_v \bar{\Q}_u  }\;.
\eeq
Note that one can get $\rho_v$ from \(\rho_u\) by exchanging indices $u\leftrightarrow v$ and $g\to 1/\bar g$.

We see that the densities and thus both $G_u$ and $G_v$ now can be expressed
solely in terms
of three polynomials $\Q_v,\;\Q_u,\;\phi$ and a function $g(x)$, generating the gauge
transformation relating the two wings. It is left only to find a closed
equation on  $g(x)$. We do it in the following subsection.

\subsubsection{Closed equation on the gauge  function $g(x) $}

In the previous subsection, we managed to express all relevant quantities
in terms of three polynomials $\Q_{v},\;\Q_u,\;\phi$ and a complex function $g(x)$. Using
the condition that two solutions of Hirota equation are related by the gauge
transformation generated by $g(x)$ we can write a closed equation on that
function. In particular, using the fact that $\Phi_v$ and $\Phi_u$ are related by  the gauge transformations (\ref{eq:GAUGE}) we obtain
\beq\la{eq:fgf}
\Phi_v=g^+g^- \Phi_u\;.
\eeq
It gives  a closed relation on $g$ which  we can rewrite, assuming that $g$ is regular within the physical strip, as follows\beq\la{gex}
g= ie^{\half i L \sinh(\pi x)} \left(-\frac{\Phi_v}{\Phi_u}\right)^{*s}\;,
\eeq
where  the zero mode of the inverted operator was chosen to  ensure the proper large $x$ asymptotic.
Indeed with this choice $T_{-1}^u=\frac{T_{-1}^v}{g^-\bar g^+}\sim e^{-L\cosh(\pi
x)}$ leading to the right behavior of $Y_0$ (see \eq{TY}) at large \(L\). Using (\ref{eq:RESOLV})  this can be re-casted as
\beq
g=i e^{\frac{1}{2} i L \sinh(\pi x)} S(x) \exp\[ s*G_v(x-i/2-i0)-s*G_u(x+i/2+i0) \] \la{eqg}
\eeq
where we used  (\ref{eq:GPhiT}) and the identity
\beq
\left(\frac{\phi^-}{\phi^+}\right)^{*s}=\left(S^+S^-\right)^{*s}= S\;,
\eeq
following from the crossing relation (\ref{crossing}). We remind that $\phi(x)=\prod_{j=1}^N (x-\theta_j)$ and $S(x)=\prod_{j=1}^N S_0(x-\theta_j)$.

  The  closed NLIE \eqref{eqg} for $g(x)$ is our main result. Together with the expressions for the densities in terms of $g$ (\ref{rhoUV}-\ref{rhoUVbar}) it allows us to calculate    $(1+Y_0)$ and thus to obtain the energy of a state  (\ref{E:eq}).

In what follows in this section we will
rewrite it in a little more convenient form which will be useful for the  numerical computations for particular states.

 Conjugating the last  equation we find
\beq
\bar g=-i e^{-i  L/2 \sinh(\pi x)} S^{-1}(x) \exp\[ s*G_v(x+i/2+i0)-s*G_u(x-i/2-i0) \]\;. \la{eqbarg}
\eeq

Finally it is useful to translate these equations into an equation for the phase $g/\bar g$
 \beq
\la{phase} \frac{g}{\bar g}= -e^{i L \sinh(\pi x)} S^2(x) \exp\left(\frac{1}{2}\[K_0^-*(\rho_u+\rho_v)-K_0^+*(\bar\rho_u+\bar\rho_v)\]\right)\;,
\eeq
where \(K_{0}=\frac{1}{2\pi i}\d_x\log S_0^2 \) and we used
\beq\la{GGKK}
G_w(x+i/2+i0)-G_w(x-i/2-i0)=K_1(x+i/2-i0)*\bar\rho_w-K_1(x-i/2+i0)*\rho_w
\eeq
and the convolution form of the dressing kernel as $K_0=2 s*K_1, K_{1}(x)=\frac{2}{\pi(4x^2+1)}$ .
For the squared norm $g \bar g$ we get
from \eq{gex}
\beq\la{eq:gbg}
g\bar g
=\left(\frac{\Phi_v}{\Phi_u}\frac{\bar\Phi_v}{\bar\Phi_u}\right)^{*s}
=\left[\frac{\Phi_v\bar\Phi_v}{T_0^{v+}T_0^{v-}}\frac{T_0^{u+}T_0^{u-}}{\Phi_u\bar\Phi_u}
\right]^{*s}\frac{T_0^{v}}{T_0^{u}}=\exp(G_v-G_u)\;,
\eeq
where we used $\frac{1}{Y_0}Y_0=1$ inside the square brackets to get the last
equality.
This equation can be also obtained from the gauge transformation $T_0^v=g\bar g T_0^u$ .
%Another useful relation is
%\com{How to make it useful?}
%
% \beq
% \frac{g^{+}}{g^{-}}= e^{-L \sinh(\pi x)} \frac{S^{+_{}}}{S^{-}} \exp s*\left[ %G^{\downarrow}_v(x)-G_v(x-i)-G_u(x+i)+G^{\uparrow}_u(x)\]
%\eeq
%
As we shell see in section \ref{sec:NUM}  these equations  can be efficiently solved numerically, by iteration, where at each iteration step a \textit{single} convolution integral arises involving the densities $\rho_u$ and $\rho_v$.

We use \eq{eq:RHOu} and \eq{eq:fgf} together with our analyticity assumptions
to constrain $g,\Phi_{u,v}$ and $T_0^{u,v}$. In the next section we will
fix the remaining finite number of complex parameters - zeros of polynomials $\Q_v,\Q_u$  and the real zeroes of $\phi$. After that one can use \eq{HSOL} to construct all $T_k$. In
 appendix C we show that all $T_k$ obtained in this way will be real functions
 and thus Hirota equation for them is satisfied. This means that we solved
 indeed
 the Y-system with the right physical analytic properties for the solutions.

\subsubsection{Finite size  Bethe equations and the energy}

Finally it is  left to explain how to fix the finite number of constants, the Bethe roots $\theta_j$, $u_j$ and $v_j$, which are  zeros of the polynomials $\phi\;,\Q_u,\;\Q_v$ and which completely characterize a state. The zeros of $\phi$ are by definition the zeros of $T_0$ which means that at these points $e^{\rho_u(\theta_{j}\pm i/2)}=e^{\rho_v(\theta_{j}\pm i/2)}=0$ as we can see from \eq{eq:RHOu}. It can be also written as follows \beq
\frac{\Q_u^+\Q_v^+}{\bar \Q_u^- \bar\Q_v^-}\frac{g}{\bar g}=1 \,\, , \,\, \quad \text{at}\;\;x=\theta_j \,, \la{baeth}
\eeq
which we can rewrite using \eq{phase} as
\beq
e^{-i L \sinh(\pi \theta_j)}=-S^2(\theta_j)\frac{\Q_u^+\Q_v^+}{\bar \Q_u^- \bar\Q_v^-}  \exp\left(\frac{1}{2}\[K_0^-*(\rho_u+\rho_v)-K_0^+*(\bar\rho_u+\bar\rho_v)\]\right) \,\, , \,\, \quad \text{at}\;\;x=\theta_j \,. \la{baeth2}
\eeq
Note that when $L\to\infty$ we can neglect the last factor to get precisely the usual infinite volume ABA \eq{BAEmiddle0}.
The equations for the auxiliary Bethe roots $u_j$ can be derived in many alternative ways. The most standard way is to demand analyticity of $T_1$
at $x=u_j$ (see \eq{JUMPS})
\beqa\la{auxu}
\Phi_u\bar \Q_u^{--}+T_0^{u-}\Q_u^{++}=0 \;\;, \;\;  \text{at}\;\; x=u_j \,.
\eeqa
We see that in general there is no reason
to assume $u_j$ to be real when $L$ is finite. Using the resolvent $G_u$ to represent $T_0$ and $\Phi_u$ appearing in this expression we get the auxiliary Bethe equations following from  (\ref{auxu}) in the form
\beq
1=- \frac{\phi^- \Q_u^{++}}{\phi^+\bar \Q_u^{--}}P(u_{j})\;, \la{baeu0}
\eeq
where $P(x)$ is defined on the upper half plane by
\beq
P(x)=\exp \[ K_1(x-i/2)*\rho_u-K_1(x+i/2)*\bar \rho_u\]\;\;,\;\;\im x>0\;, \la{Px}
\eeq In the large $L$ limit $P(x)\sim 1$ and we get the ABA \eq{BAEuv0}.
A similar equation fixes the roots \(v_j\).

The integral equation (\ref{eqg}) together with the equations (\ref{baeth2}), (\ref{baeu0}) fixing the zeros of the polynomials $\Q_u,\;\Q_v,\;\phi$  are the complete set of equations which one should solve to find the full spectrum of the $SU(2)\times SU(2)$ principal chiral field. Once $g(x)$ and the positions of the zeros $\theta_j,u_j,v_j$ are found, we can compute the exact energy of the corresponding quantum state from  \eqs{E:eq}{Y0RHO} and \eqref{eq:RHOu}
\beq\label{ELrho}
E=\sum_{k=1}^{N}\cosh(\pi\theta_{k})\,-\frac{1}{2}\,\int \cosh(\pi x)(\rho_u+\bar \rho_u)dx\;.
\eeq
where we can use due to \eqref{eq:RHOu} \(\rho_u+\bar \rho_u=\rho_v+\bar \rho_v=\log\left| \frac{\frac{g^{+}}{\bar g^+}\Q^{++}_u \Q_v^{++}- \bar \Q_u\bar\Q_v}{\frac{g^{+}}{g^-}\Q_u^{++}\bar \Q_v^{--}-\Q_u\bar\Q_v}\right|\).

Let us remark that our construction for a general state in this paper was based on the assumption that in the asymptotic regime \(L\to\infty\) all the roots \(u_j,v_j\)   become real.However, it is well known that the complex solutions are also possible.  We hope that even in this case  our  equations  maintain their form, although this situation deserves a special care.

In section 4 we will explain how to efficiently implement these equations for numerical study. Before that, in the next section we will study the large $L$ behavior of these equations thus reproducing not only the large volume results of section \ref{ABA} but also the subleading  corrections (L\"uscher corrections).

%%%%%%%%%%%%%%%%%%%%%%%%%%%%%%%%%%%%%%%%%%

\subsection{Large volume limit: ABA and L\"uscher corrections} \la{Luscher:sec}

The $SU(2)$ principal chiral field spectrum is given by \eqref{E:eq}, or \eqref{ELrho}.
As we have seen in the previous section, in the large $L$ limit the Bethe roots $\theta_j$ are given by their asymptotic values obtained from a solution to the asymptotic Bethe equations, and since $Y_0$ and $\rho$'s are exponentially small we can drop the integral contribution in  \eqref{baeth2} and \eqref{baeu0} and recover the usual asymptotic spectrum. In this section we focus on the leading finite size corrections to this result.

Due to these corrections auxiliary roots
$u_j$ and $v_j$  become complex even if they were real asymptotically at large $L$.
In this section we denote the real part of the roots $u_j, v_j$  by  $U_j$ and $V_{j}$ while the (small) imaginary parts we denote by $\Delta u_j$ and $\Delta v_j$. The positions of the momentum carrying roots $\theta_j$ are
 also corrected, however they stay real. We will also use the notation
\beq
\QQ_u(x)=\prod_j{(x-U_j)}\;\;,\;\;\QQ_v(x)=\prod_j{(x-V_j)}\;.
\eeq
We have to compute the first correction to the positions of the Bethe roots. To the leading order we can drop the exponentially small densities $\rho_u$ and $\rho_v$ in \eq{eqg}, to get \beq
g(x)\simeq i S(x) e^{i L/2 \sinh(\pi x)}\;\label{leadingg}
\eeq
which we can use to compute the spectral densities from (\ref{rhoUV}-\ref{rhoUVbar}).
%\beq
%g^{+}=iS^+e^{-1/2L\cosh(\pi x)}\;\;,\;\;g^{-}=iS^-e^{+1/2L\cosh(\pi x)}
%,\quad\bar g^+=-i\frac{1}{S^+}e^{+1/2L\cosh(\pi x)}\;\;\eeq
We see that some terms in the  expression for $\rho_u$ are exponentially suppressed and we can expand
\begin{equation}\label{eq:RhouCorr}
\rho_{u}\simeq \frac{\bar\Q_u-\Q_u}{\Q_u}+
e^{-L\cosh(\pi x)}
\frac{S^+}{S^-}
\frac{\Q_u^{++}}{\phi^+}\frac{\Q_v^{++}\phi^-+\Q_v^{--}\phi^+}{\Q_v \Q_u}\simeq
\frac{\bar\Q_u-\Q_u}{\QQ_u}+
\frac{\QQ_u^{++}T_{-1}^u}{\QQ_u\phi^+}\;.
\end{equation}
In the last equality we neglect the small imaginary part of the axillary
roots and
we  use \eqref{JUMPS} to the leading order together with the gauge transformation $T_{-1}^v=g^-\bar g^+ T_{-1}^u$ and the crossing relation \(S^+S^-=\frac{\phi^-}{\phi^+}\).

The poles at $x=U_j$ should cancel, due to  \eq{baeu0}, among the first and the
second term since the density by definition  is regular. We introduce the notations $\rho_u^{(1)}$ and $\rho_u^{(2)}$ for the first and the second term in \eqref{eq:RhouCorr}. The first one can be simply written as
\beq
\rho^{(1)}_{u}\simeq \sum_j\frac{2\Delta u_j}{x-U_j}\;.\la{rh1}
\eeq
Since the whole density is regular we can apply the principal part prescription to the finite integrals in (\ref{phase}) without changing the result. Having done so we are free to split the convolutions into convolutions with $\rho^{(1)}$ and $\rho^{(2)}$. In (\ref{baeth2}) we should then expand the factor
\beq
\left[\frac{\Q_u^+}{\bar \Q_u^-}\exp\left( i \im
\dashint K_0^-(x-y) \rho_u^{(1)}(y)  \right) \right]\exp\left(i \im
\dashint K_0^-(x-y) \rho_u^{(2)}(y)  \right) \,,
\eeq
 and the similar factor for the $v$ roots, to the next to leading order. We notice that $\rho_u^{(1)}$  is purely imaginary to the leading order, as seen from  \eq{rh1},
and therefore we can simplify the term in the square brackets\small\textsl{}
\beq
\frac{\Q_u^+}{\bar \Q_u^-}\exp\left( \frac{1}{2} (K_0^-+K_0^+)* \rho_u^{(1)}  \right)\simeq\frac{\Q_u^+}{\bar \Q_u^-}\left(1+ K_1* \rho_u^{(1)}  \right)
 =\frac{\Q_u^+}{\bar \Q_u^-}\left(1+ \frac{\rho_u^{(1)+}+\rho_u^{(1)-}}{2}   \right)\simeq\frac{\QQ_u^{+}}{\QQ_u^{-}}\; \la{derivation}
\eeq
where the convolutions are understood in the sense of principal value.
Thus in the Bethe equations (\ref{baeth})  in this approximation the imaginary parts of the
axillary roots cancel against the contribution from $\rho^{(1)}$ and we simply
get
 \begin{equation}
-e^{i L \sinh(\pi x)} S^2 \frac{\QQ_u^{+}\QQ_v^{+}}{\QQ_u^{-}\QQ_v^{-}}  = \exp\left(-i \im
K_0^-* \left[\rho_u^{(2)}+\rho_v^{(2)}\right]  \right) \;\;\text{at}\;\; x=\theta_j\;. \la{corrTh}
\end{equation}
Proceeding in the same fashion in the \eq{baeu0} for the auxiliary Bethe roots  we arrive at a similar conclusion. Namely only the real parts of the auxiliary
roots survive when we separate the density into $\rho^{(1)}$ and $\rho^{(2)}$
\beq
-\frac{\phi^-}{\phi^+}\frac{ \QQ_u^{++}}{\QQ_u^{--}}=\exp\left(-2i\IM K_{1}^{-}*\rho_{u}^{(2)}\right)\;\;\text{at}\;\;x=U_j\;. \la{corrU}
\eeq
See appendix \ref{appE} for details.
%To see this we use that $\rho^{(1)}$ is purely imaginary, to get\footnote{See the definitions of the kernels \(K_n\) in the Appendix \ref{AppA}}
%\beq
%P(x)=\exp\left( K_{2}*\rho^{(1)}+(K_1^{-}*\rho^{(2)}-c.c.)+\frac{Y_0(x)}{2}\right)
%\eeq
%where the convolutions are understood in the sense of principal value
%as before. Moreover we notice that
%\beqa
%\frac{ \Q_u^{++}(u_i)}{\bar \Q_u^{--}(u_i)}\exp\left(K_2*\rho^{(1)}(u_i)\right)
%&=&\frac{ \QQ_u^{++}(u_i)}{ \QQ_u^{--}(u_i)}\\
% \frac{\phi^-(u_i)}{\phi^+(u_i)}\frac{ \QQ_u^{++}(u_i)}{\QQ_u^{--}(u_i)}\exp\left(\frac{Y_0(u_i)}{2}\right)
% &=&
%  \frac{\phi^-(U_i)}{\phi^+(U_i)}\frac{ \QQ_u^{++}(U_i)}{\QQ_u^{--}(U_i)}\;.
%\eeqa
We see that all terms except for the convolutions with $\rho^{(2)}$ have a simple effect of absorbing the  imaginary parts of the Bethe roots.

It turns out that the remaining convolutions, containing $\rho^{(2)}$, can be nicely written in terms of the leading order $Y_0$ found before in (\ref{LargeLY}),
\begin{equation}\la{LargeLY2}
Y_0(x) \simeq e^{-L\cosh(\pi x)} \left(\frac{\Q_u^{++} \phi^-+\Q_u^{--}\phi^+}{\Q_u}\right)\left(\frac{\Q_v^{++} \phi^-+\Q_v^{--}\phi^+}{\Q_v}\right)\frac{(S^+)^2}{(\phi^-)^2}\;,
\end{equation}
where the $\theta_j$ appear in $\phi$ and $S$ while the $u_j$ ($v_j$) auxiliary roots appear in the corresponding Baxter polynomials $\Q_u$ ($\Q_v$). Notice that this quantity is already exponentially small, so  we can take here the asymptotic values for the auxiliary roots. As explained in detail in appendix \ref{appE}, the quantities inside the principal part integrals are related to the derivative of this function with respect to $\theta_k$ or $u_k$ and $v_k$ which we treat in (\ref{LargeLY2}) as independent variables.  More precisely we have the remarkable identities
\beqa
i \, \im  \left( K_0^-(\theta_i-y)\left[ \rho_u^{(2)}(y)+\rho_v^{(2)}(y)\right] \right)&=&-\frac{\partial_{\theta_i} Y_0(y)}{2\pi i} \;\;,\\   \;\;
2i \, \im \left(K_1^{-}(u_j-y)\rho_u^{(2)}(y)\right)&=&+\frac{\partial_{u_{i}} Y_0(y)}{2\pi i} \,. \la{remark}
\eeqa
Thus we finally obtain the corrected Bethe ansatz equations in the following elegant form:
\beqa
-\frac{\phi^+}{\phi^-}\frac{ \QQ_u^{--}}{\QQ_u^{++}}&=&\exp\left(\dashint\frac{\partial_{U_{j}} Y_0(y)}{2\pi i}dy\right)\;\;\text{at}\;\;x=U_j\;\nn,\\
-e^{i L \sinh(\pi x)} S^2 \frac{\QQ_u^{+}\QQ_v^{+}}{\QQ_u^{-}\QQ_v^{-}}  &=&\exp\left(\dashint\frac{\partial_{\theta_j} Y_0(y)}{2\pi i}dy \right) \;\;\text{at}\;\; x=\theta_j\;,\label{ourL}\\
-\frac{\phi^+}{\phi^-}\frac{ \QQ_v^{--}}{\QQ_v^{++}}&=&\exp\left(\dashint\frac{\partial_{V_{j}} Y_0(y)}{2\pi i}dy\right)\;\;\text{at}\;\;x=V_j\;.\nn
\eeqa
It is not completely surprising that we managed to express everything in
terms of $Y_0$. To the leading order,  $Y_0$  can be expressed in terms of S-matrix only: it is the relevant eigenvalue of
the operator\begin{equation}
e^{- L \cosh(\pi x)} {\rm tr}\left( \hat S_{01}(x-\theta_1) \hat S_{02}(x-\theta_2)\dots  \hat S_{0N}(x-\theta_N) \right) \la{Y0leading}
.\end{equation}
We see that (\ref{ourL}) corresponds precisely to the conjectured equation (27) in \cite{BJ} only inside the $U(1)$ sector. However, our result is different from  outside
the $U(1)$ sector when there are axillary roots $U_j$ and $V_j$.
Finally the equation for the energy of the state corrected by the finite
size effects is given by \eq{E:eq} in terms of $Y_0$ in the leading approximation and the corrected positions
of the roots $\theta_j$ which  should be found from \eq{ourL}.

 The right-hand sides of the corrected Bethe equations \eqref{ourL} have a simple interpretation: for the middle equation, it reflects the contribution of scattering of the ``physical" particles off the virtual ones on the cylinder, whether as the other two reflect the same effect for the ``magnons" responsible for the isotopic degrees of freedom of the particles.    Although these equations are derived here only for a particular model their form looks very universal and can be  immediately generalized to any other integrable sigma model where the exact scattering matrix is known.

\subsubsection{Single particle case}
In this section we  consider the single particle case for the L\"uscher-type correction of the previous subsections. This analysis was done in a more general context in \cite{BJ}.

When we have a single particle with momentum $\theta_1$ (\ref{ourL}) yields simply
\begin{equation}
 L\sinh \left(\pi \theta_1 \right)  =2\pi n -\dashint \frac{dy}{2\pi } \partial_{\theta_1} Y_0(y) \la{corrected1} \,,
\end{equation}
which corrects the leading order quantization condition
\beq
L \sinh(\pi \theta_1^0)=2\pi n \,.
\eeq
Now, from (\ref{LargeLY2}) we see that the $x$ dependence in $Y_0(x)$ comes from the exponential factor $e^{-L \pi \cosh(\pi x)}$ and also from the combinations $x-\theta_j,x-u_j$ and $x-v_j$ appearing in the remaining terms in this expression. Thus
\begin{equation}
\partial_y Y_0(y)=- L \pi \sinh(\pi y) Y_0(y)-\sum_{k=1}^N \partial_{\theta_k} Y_0(y)-\sum_{k=1}^{J_u} \partial_{u_k} Y_0(y)-\sum_{k=1}^{J_v} \partial_{v_k} Y_0(y) \,,
\end{equation}
which, in the case we are considering, with $N=1$ and $J_u=J_v=0$,  allows us to simplify (\ref{corrected1}) to
\begin{equation}
 L\sinh \left(\pi \theta_1 \right)  =2\pi n+ \int \frac{dy}{2 }  L \sinh(\pi y) Y_0(y) \,,
\end{equation}
so that the leading finite size correction to the energy (\ref{E:eq}) reads
\begin{equation}
E(L)-   \cosh(\pi \theta_1^0) \simeq -\frac{1}{2}\int \cosh(\pi y) \left(1- \tanh(\pi y)\tanh(\pi \theta_1^0)\right)e^{- L \cosh(\pi y)} {\rm tr}  \, \hat S_{01} (y-\theta_1^0) \,,
\end{equation}
precisely as expected for the L\"uscher corrections \cite{Luscher:1985dn}.

\section{$SU(2)$ Chiral Gross-Neveu model and related models} \la{sec:GN}
Our NLIE resembles the Destri-deVega equation and, at least in the  cases the last one is known, can even coincide with it.
In the cases when the DdV equation is not known, like the  \(SU(2)_{L}\times SU(2)_{R} \) PCF, or O(4) model  studied in this paper, we obtain a new, DdV-like equation. In this  subsection,  to demonstrate our method, we show how to reproduce the DdV equation  for  the chiral  $SU(2)$ Gross-Neveu model on a finite circle.

 The TBA equations for this model are given by the same $Y$-system (\ref{Y:eq}) with an important difference that
$Y_{s<0}=0$ (see Fig.\ref{diagrams}). In particular, since $Y_{-1}=0$ we have $T_{-2}=0$ which implies due to the \eqs{LAX}{eq:Hir}
% yields
\begin{eqnarray}\label{laxGN}
0=T^{--}_{-1}\bar Q -
\bar \Phi  Q^{--}\;\;,\;\;
T_{-1}^+T_{-1}^-=\Phi^-\bar\Phi^+\;.
\end{eqnarray}
Then it is immediate to check that the quantity

%so that if we define
\beq
\AAA\equiv \frac{Q^{+}}{\bar Q^-}\frac{T_{-1}^-}{\Phi^-}
=\frac{Q^{+}}{\bar Q^-}\frac{\bar\Phi^+}{T_{-1}^+}%=\frac{1}{\bar A}
\eeq
%and therefore
%\begin{eqnarray}\label{laxGN}
%T_{-1}^+ T_{-1}^-=\Phi^-\bar \Phi^+
%%\frac{\Phi\left(x-i/2 \right)\bar Q\left(x+i/2\right)}{Q(x-i/2)}\frac{\bar \Phi\left(x+i/2 \right)Q\left(x-i/2\right)}{\bar Q(x+i/2)}
%\end{eqnarray}
satisfies, due to \eq{laxGN},
the relation\beq
\AAA^+\AAA^-=\left(\frac{Q^+}{\bar Q^-}\right)^+\left(\frac{Q^+}{\bar Q^-}\right)^- \frac{\bar \Phi}{\Phi} \,.
\eeq
Note that $\AAA$ is a pure phase on the real axe. Thus, restoring the proper zero mode exponential, we find
\beq
\AAA=e^{iL\sinh(\pi x)}\frac{Q^{+}}{\bar
Q^{-}}\left(\frac{\bar\Phi}{\Phi}\right)^{*s} \,. \la{intA}
\eeq
As before, to make this a closed equation on $\AAA$ we introduce the resolvent $G$ and the density $\rho$
\begin{equation}\la{eq:RESOLVGN}
G(x)=\frac{1}{2\pi i}\int
\frac{\rho(y)}{x-y-i/2}dy-\frac{1}{2\pi i}\int
\frac{\bar\rho(y)}{x-y+i/2}dy\;\;,\quad\;\; \rho=\log\frac{T_0^+}{\Phi} \,.
\end{equation}
Again analytic properties of $T_0$ and $\Phi$ lead to \eq{eq:GPhiT}. Using the linear problem (\ref{LAX})  we can write $T_{-1} Q^{++}-T_0^+Q=-\Phi\bar
Q$ and we see that
\beq\label{DENGN}
\rho=\log\[\frac{\bar Q}{Q}\left(1+\AAA^+\right)\]\,.
\eeq
Thus \(\AAA\) satisfies the closed equation  for $\AAA$ since $\Phi=G(x+i/2+i0)$.  We see that $\AAA(x)$ plays a similar role
as $g(x)$ in PCF.
We can easily compute $Y_0$ in terms of $\AAA$
\beq
1+Y_0=e^{\rho}e^{\bar\rho}=\left(1+\AAA^+\right)\left(1+1/\AAA^-\right)\;\la{eq2}.
\eeq
We see that the factors of $Q$  cancel from this expression. It is also possible to
write \eq{intA} in a simpler form
without factors of $Q$ using the useful identities. First using \eq{GGKK}
and assuming that the  density is regular not only on the real axes but also for \(-1/2<\im x\leq0\) we write
\beq\label{MainEq}
\left(\frac{\bar\Phi}{\Phi}\right)^{*s}=S_{\theta}\exp\left(\frac{1}{2}K_0^-*\rho-\frac{1}{2}K_0^+*\bar\rho\right)
=S_{\theta}\exp\left(\frac{1}{2}K_0*[\rho^{-_{+}}-\bar\rho^{+_-}]\right)\;,
\eeq
where $\rho^{-_+}\equiv\rho(x-i/2+i0)$. Using the following relation
\beq
\frac{1}{2}K_0*\log\[\frac{\bar Q^+ \bar Q^-}{Q^- Q^+}\]=
\log\[\frac{S_{u}^+}{S_{\bar u}^+}\frac{\bar Q^-}{Q^-}\]\;,
\eeq
where
\beq
S_{\theta}(x)=\prod_{j=1}^N S_0(x-\theta_j)\,\,,\,\,S_{u}(x)=\prod_{j=1}^{J_u} S_0(x-u_j)\,\,,\,\,S_{\bar u}(x)=\prod_{j=1}^{J_u} S_0(x-\bar u_j)\; \la{eq1}
\eeq
 we can  get rid of $Q$'s in \eqref{MainEq}  and  finally obtain the
%and thus we are left with
%\beq
%A=e^{iL\sinh(\pi x)}S
%\frac{Q^{+}}{\bar Q^{-}}\exp\left(i\im K_0^-*\log\[\frac{\bar Q}{Q}\left(1+A^+\right)\] \right)
%\eeq
%which can be brought to the known
known DdV equation
\beqa
\AAA%&=&e^{iL\sinh(\pi x)}S_\theta
%\frac{Q^{+}}{\bar Q^{-}}\exp\left(i\im K_0*\log\[\frac{\bar Q^-}{Q^-}\left(1+A_{+i0}\right)\]\right)\\
%&=&e^{iL\sinh(\pi x)}S_\theta
%\frac{Q^{+}}{\bar Q^{-}}\exp\left(\frac{1}{2}K_0*\log\[\frac{\bar Q^+ \bar Q^-}{Q^- Q^+}\frac{1+A_{+0}}{1+1/A_{-0}}\]\right)\;,
%&=&e^{iL\sinh(\pi x)}S_\theta
%\frac{S_{u}^+}{S_{\bar u}^+}\frac{Q^+}{Q^-}\exp\left(i\im K_0*\log\[1+A_{+i0}\]\right)
&=&e^{iL\sinh(\pi x)}
\frac{S_\theta}{S_{\bar u}^+S_{u}^-}\exp\left({i\im K_0*\log\[1+\AAA(x+i0)\]}\right)\;.
\eeqa
 Then for $Y_0$ we get the standard relation
which should be used to compute the exact spectrum from

\begin{equation}\la{eq3}
E(L)=-\frac{1}{2}\int m\cosh(\pi\theta)\log(1+ Y_0)+\sum m\cosh(\pi \theta_j)\;.
\end{equation}
Furthermore, both BAE -- for the physical rapidities $\theta_j$ and for the magnon rapidities $u_j$ -- can be written
\beq
\AAA(\theta_j)=-1\;\;,\;\;\AAA^+(u_j)=-1\;. \la{eq4}
\eeq
We also notice that $\AAA(x)$ has poles at $x=\bar u_j+i/2$.
To see that we use
\beqa
&&T_0(\theta_j)=0\;\;\Rightarrow\;\;1+\AAA(\theta_j)=\frac{Q^-(\theta_j)}{\bar Q^-(\theta_j)}\frac{T_0(\theta_j)}{\Phi^-(\theta_j)}=0\,,\\
&&\Phi(u_j)\sim 1\;\;\Rightarrow\;\;1+\AAA^+(u_j)=\frac{Q (u_j)}{\bar
Q(u_j)}\frac{T_0^+(u_j)}{\Phi(u_j)}=0\,.
\eeqa
Equations (\ref{eq1}),(\ref{eq2}),(\ref{eq3}) and (\ref{eq4}) are precisely the DdV equations derived in \cite{Destri:1987ug,Fioravanti:1996rz}\footnote{In these papers the sine-Gordon model was considered. The $SU(2)$ Chiral Gross-Neveu model is a simple limiting case of this theory, see discussion in the paragraph below.}! As shown in this section our method can  be directly generalized to other models whose TBA $Y$-system equations are known. It would be very interesting to make a systematic study of such models using our formalism.

For example, a simple generalization of the case studied in this section is obtained by considering the functions $T_k$, $\Phi$ as well as the Baxter functions $Q$ to be periodic in the imaginary directions with period $i\nu$. This amounts to considering the trigonometric solutions of Hirota equation (\ref{eq:Hir}) and the corresponding linear problem (\ref{LAX}) -- this should correspond to the sine-Gordon model \cite{Balog:2003xd}. We take the Baxter polynomials $Q(x)$ and the large $L$ limit of $T_0,\Phi^-,\bar \Phi\simeq \tilde\phi$ to be
\beq
\tilde Q(x)=\prod_{j=1}^J\frac{\sinh\frac{\pi}{\nu}(x-u_j)}{\sinh\frac{\pi}{\nu}}\,\,,\,\, \tilde \phi(x)=\prod_{j=1}^N\frac{\sinh\frac{\pi}{\nu}(x-\theta_j)}{\sinh\frac{\pi}{\nu}} \,
\eeq
instead of polynomials. Then most of the formulae in this section go through with minor modifications. For example, instead of the $SU(2)$ Chiral Gross-Neveu S-matrix $S_0=\left(\frac{x-i/2}{x+i/2}\right)^{*s}$ we will find the sine-Gordon dressing factor
$$
\tilde S_0=\left(\frac{\sinh\frac{\pi}{\nu}(x-i/2)}{\sinh\frac{\pi}{\nu}(x+i/2)}\right)^{*s}=-i \exp \int_0^{\infty} \frac{\sin(\omega x)}{\omega} \frac{\sinh\left(\frac{\nu-1}{2}\omega\right)}{\cosh\left(\frac{\omega}{2}\right)\sinh\left(\frac{\nu}{2}\omega\right)}\,.
$$
Thus it seems that our method  allows  to derive the sine-Gordon DdV equations of \cite{Destri:1987ug,Fioravanti:1996rz} in an easy way. For $\nu\to\infty$ we recover the $SU(2)$ chiral Gross-Neveu model. For an integer $\nu$ the Y-system can be truncated as represented in figure \ref{diagrams}, see e.g. \cite{Balog:2003xd}.

Another interesting class of models which one could analyze using our formalism is represented  by the so called sausage model (see e.g. \cite{Fateev:1996ea}). This model can be considered as a generalization of  the $O(4)$ model, or \(SU(2) \) PCF, in the same sense as the sine-Gordon model is a generalization of the $SU(2)$ chiral Gross-Neveu.  The inhomogeneous XXX-spin chain present in the $SU(2)$ Gross-Neveu model and describing the isotopic degrees of freedom is generalized in  sine-Gordon model    to the XXZ chain, with the anisotropy  parameter $\nu$ introduced above. Similarly, the sausage models can be seen as two interacting inhomogeneous XXZ chains parameterized by the inhomogeneities $\theta_1,\dots, \theta_N$ and anisotropies $\nu$ and $\nu'$. It would be very interesting to generalize our \(O(4)\) model  results  to this more general class of models.

Our approach to deriving DdV like equations is strongly based on a smooth interpolation starting from the IR asymptotic Bethe ansatz description; hence, by construction, our states are very well identified. On the other hand we did not carry out a detailed study of complex solutions such as the states represented by Bethe strings in the large $L$ limit; for these states some of our formulae might need to be modified. Within the DdV approach based on  descritizations of integrable models, many interesting complex solutions were studied: e.g. holes, special objects, wide roots, self-conjugate roots etc. It would be interesting to complete our approach to include all physical complex solutions and thus obtain a precise dictionary between these two approaches. In particular this would teach us which solutions to DdV NLIE  correspond to physical states.

\section{Numerics}\label{sec:NUM}

In this section we explain how to efficiently solve numerically the equations derived in the Sec.\ref{GEN}.

\subsection{Implementation of numerics and Mathematica code}
For simplicity let us focus on the  $U(1)$ sector where we consider $M$--particle quantum states %\footnote{Here we se $M$ instead of $N$ to denote the number of particles because we below we will present the Mathematica code necessary to solve this problem and the letter $N$ is reserved in Mathematica}
 with  $M$  spins pointing in the same direction in \(SU(2)_L\) and \(SU(2)_R\). Then our equations simplify considerably as was explained in subsection \ref{subsec:U1}. First of all, since there are no spins excited we have trivial Baxter polynomials $\mathcal{Q}_u=\mathcal{Q}_v=1$. Thus, from (\ref{eq:RHOu}) we see that $\rho_u=\rho_v=\rho$ with
\beq
\rho=\log\frac{\frac{g^+}{\bar g^+}-1}{\frac{g^+}{g^-}-1} \,.
\eeq
We also notice that since this is a symmetric configuration where the $u$ and $v$ root configurations are the same (there are no roots at all) we have, see e.g. (\ref{eq:gbg}), $g\bar g=1$ and thus $g(x)$ is a pure phase. In particular, for real $x$, we can simplify the density to
\beq
\rho(x)=\log\frac{\left(g^+\right)^2-1}{ | g^+ |^2-1} \,,
\eeq
from where we see that we can express it solely in terms of $g^+$. Since $g$ is a pure phase we need only to determine its argument from (\ref{phase}) which now reads
 \beq
g^2= -e^{i L \sinh(\pi x)} S^2(x) \exp \left(2i\; \im\! \[K_0^-*\log\frac{\left(g^+\right)^2-1}{ | g^+ |^2-1} \] \right)\;.
\eeq
This is almost perfect for numerical implementation but still needs to be slightly improved. The reason is that we want to iterate this equation by evaluating the right hand side for real $x$. But this will yield the updated values of $g(x)$ in the left hand side whereas for the next iteration we would need $g^+(x)$. To fix it, we simply shift $x \to x+i/2$ in this equation and define $A(x)\equiv \left(g^+(x)\right)^2$ to get
 \beq
A= - e^{- L \cosh(\pi x)} \prod_{j=1}^M S^2_0(x-\theta_j+i/2) \exp \left( K_0*\log\frac{A-1}{ |A|-1}-K_0^{++}*\log\frac{\bar A-1}{ |A|-1}- \log\frac{\bar A-1}{ |A|-1} \right)\;,
\eeq
where the convolution of $K_0^{++}$ is understood in the principal part sense. We have explicitly written $S(x)$ to render the presence of the Bethe rapidities more explicit. These are fixed by the main Bethe equation (\ref{baeth}) which in our notations is simply
 \beq
 -e^{i L \sinh(\pi x)}  \prod_{j=1}^M S^2_0(x-\theta_j)\exp \left(2i \; \im\! \[K_0^-*\log\frac{A-1}{ |A|-1}\] \right)=1 \,\, , \,\, x= \theta_i \,. \la{BAEnum}
\eeq
For completeness let us present here the  Mathematica code to  solve these two equations by iterations\footnote{One can copy the code directly to {\it Mathematica}
from .pdf}. It is a slightly simplified, and thus  less efficient, version of the code we used for the plots in figure \ref{pic4}.

First we introduce the S-matrix $S_0$ and the kernel $K_0$,
\\
{\footnotesize
\verb"          "\\
\verb"  S0[x_]=I*Gamma[-(x/(2I))]Gamma[1/2+x/(2I)]/(Gamma[x/(2I)]Gamma[1/2-x/(2I)]);"\\
\verb"  K0[x_]=D[Log[S0[x]^2],x]/(2*Pi*I);"
\verb"          "\\
}\\
Next we specify the  size $L$  and  the mode numbers $n=\{n_1,\dots,n_{M}\}$. For example, if we want to study the system with $L=1/2$ and three particles with zero mode numbers we write\\
{\footnotesize
\verb"          "\\
\verb"  n = {0, 0, 0};"\\
\verb"  M = Length[n];"\\
\verb"  L = 1/2;"
\verb"          "\\
}\\
We will perform several integrals from $-\infty$ to $+\infty$ but the integrands
have exponential tails so that it is quite useful to introduce a cut-off $X$ for all the integration intervals at this point. A reasonable cut-off is given by $e^{-L \cosh(\pi X)}=10^{-8}$. Furthermore, at each iteration step we will have to construct an updated function $A(x)$ which we do by means of an interpolation function,
\\{\footnotesize
\verb"          "\\
\verb"  X=ArcCosh[8Log[10]/L]/\[Pi];" \\
\verb"  F[S_]:=FunctionInterpolation[S,{x,-X,X},InterpolationPoints->30];"
\verb"          "\\
}\\
Next we introduce\\
{\footnotesize
\verb"          "\\
\verb"  eq[i_,v_]:=L*Sinh[Pi*x[i]]+Sum[If[i==j,0,Log[S0[x[i]-x[j]]^2]/I],{j,M}]-2n[[i]]Pi+v[[i]];"\\
\verb"  BAE[v_]:=Table[x[i],{i,M}]/.FindRoot[Table[Re[eq[j,v]],{j,M}],"\\
\verb"    Table[{x[i],2i/M-1/2},{i,M}]];"\\
\verb"          "\\
}
which yields the solution to Bethe equations $L\sinh(\pi \theta_i)+\sum_{j\neq i} \frac{1}{i}\log S_0^2(\theta_i-\theta_j)-2\pi n_i+v_i$ where $v_i$ is a perturbation to the equation number $i$. Comparing with (\ref{BAEnum}) we see that this perturbation at step $k$ will be given by the convolution appearing in (\ref{BAEnum}) evaluated at the solution $\theta_i$ computed in the previous step. The leading order BAE's correspond to $v_i=0$ and are thus given by
\\
{\footnotesize
\verb"          "\\
%&&\verb"Clear[A,\[Theta]]"\\
\verb"  \[Theta][0]=BAE[Table[0,{j,M}]]"
\verb"          "\\
}
\\
Also to leading order the function $A(x)$ will be simply given by
\\
{\footnotesize
\verb"          "\\
\verb"  A[0]=F[-Exp[-L*Cosh[Pi*x]]*Product[S0[x-\[Theta][0][[j]]+I/2]^2,{j,M}]]"
\verb"          "\\
}
\\
Then we introduce the density $\rho$ and its conjugate $\bar \rho$ at the $k$-th iteration step as
\\
{\footnotesize
\verb"          "\\
\verb"  r[k_,y_]:=Log[(A[k][y]-1)/(Abs[A[k][y]]-1)];"\\
\verb"  rc[k_,y_]:=Conjugate[r[k,y]];"
\verb"          "\\
}
\\
Finally the code
\\
{\footnotesize
\verb"     "\\
\verb"  A[k_]:=A[k]=F[-Exp[-L*Cosh[Pi*x]]Product[S0[x-\[Theta][k-1][[j]]+I/2]^2,{j,M}]"\\
\verb"  Exp[NIntegrate[K0[x-y]r[k-1,y]-K0[x-y+I]*rc[k-1,y]+1,{y,-X,x,X},"\\
\verb"    Method->PrincipalValue]-2X-rc[k-1,x]]];"\\
\verb"  phase[k_][x_]:=NIntegrate[2Im[K0[x-y-I/2]r[k-1,y]]+1,{y,-X,X}]-2X;"\\
\verb"  \[Theta][k_]:=\[Theta][k]=BAE[Table[phase[k][\[Theta][k-1][[j]]],{j,M}]];"\\
\verb"     "\\
}
yields the $k$-th iteration quantities in terms of those computed at the $(k-1)$-th step. The energy of the state is then given by\\
{\footnotesize
\verb"          "\\
\verb"  En[k_]:=Sum[Cosh[Pi\[Theta][k][[j]]],{j, M}]-NIntegrate[Re[r[k,y]]Cosh[Pi*y],{y,-X,X}];"\\
\verb"          "\\
}
For example, to obtain the result of the first $8$ iterations we simply run
\\
{\footnotesize
\verb"          "\\
\verb"  Table[En[k], {k, 0, 8}]"
\verb"          "\\
}
\\
to get \{10.2414,10.2425,10.2425,10.2424,10.2424,10.2424,10.2424,10.2424\} where we notice that the iterations are clearly converging to the exact value $10.2424$ up to the precision we are working at. It is instructive to compare these results to the value predicted by the asymptotic Bethe equations alone,
\begin{eqnarray*}
E_{\text{BAE}}=10.3388
\end{eqnarray*}
and to that predicted by the generalized Luscher formulae discussed in section \ref{Luscher:sec} which gives
\begin{eqnarray*}
E_{\text{L\"uscher}}=10.2396
\end{eqnarray*}
%In figure \ref{Luscher:fig} we plot the comparison between the exact numerical %result and the predictions from the generalized L\"uscher treatment for %several values of $L$ for the two particle state $\theta_{00}$.

\begin{figure}[t]
\epsfxsize=10cm
\centerline{\epsfbox{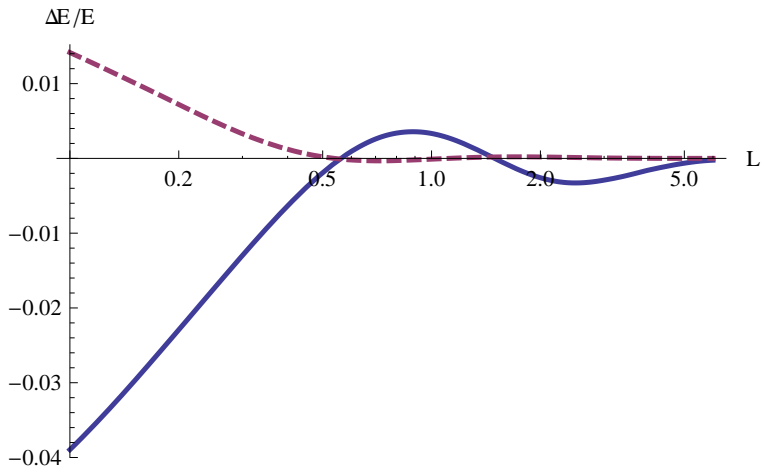}}
\caption{\label{Luscher:fig}
\small\textsl{Difference between the predictions from the asymptotic Bethe ansatz and from the generalized L\"uscher formulas to the exact (numerical) results for the two particle state polarized in both $SU(2)$ [$u,v,\theta_{-11}$]. The solid (blue) line represents $\frac{E_{exact}-E_{ABA}}{E_{exact}}$ while the dashed (red) curve depicts $\frac{E_{exact}-E_{Luscher}}{E_{exact}}$. It is clear that the latter approximates the exact results with excellent accuracy, especially for large $L$ of course.
}
}
\end{figure}

\subsection{Discussion of numerical results}\label{NUMexpl}

Now we will try to interpret the behavior of various states on the fig.\ref{pic2}  as functions of the volume \(L\). Let us start from the vacuum, the lowest plot there. At very small \(L\), the \(O(4)\) model should become a 2d CFT of three massless bosons: if we introduce   in \eqref{O4sigma}  a rescaled field \(\vec Y=e^{-1}_{0}(L)\left( X_{1},X_{2},X_{3} \right)\) and
\(X_4=\sqrt{1-e^2_0(L)(Y_1^2+Y_{2}^{2}+Y_{3}^{2})}\),  where \(e^{2}_{0}(L)\simeq\frac{2\pi}{|\log L|} \) is the effective charge,  very small in this limit (the effective radius of the \(S^{3} \) sphere \(R(L)=e_{0}^{-1}(L)\)  is very big), the action will be

\begin{equation}\label{O4sigma}
\mathcal{S_{\sigma}}_\text{} = \int dt \,dx \:\ \sum_{a=1}^{3}(\partial_{\alpha} Y_{a}^{_{}})^{2}\quad +\,O\left(e^{4}_{0}(L)\right).
 \end{equation}

In the ground state, the Casimir effect will define the limiting energy: \(E_{0}\simeq-\frac{\pi c}{6L}+O(1/\log ^{4}(L^{-1}))\), with the central charge \(c=3\), which gives  \(E_{0}\frac{L}{2\pi}\simeq-\frac{1}{4}\), the value compatible with \(-0.18\) of the fig.\ref{pic2}.
\footnote{The convergence to the limiting value is very slow at \(L\rightarrow 0\). At \(L=0.1\) for our calculations we are still far from the limiting value of the energy. In \cite{Balog:2003yr}, where the numerics reached \(L=10^{-6}   \), the result is \(-0.226 \), considerably closer to the limiting value. }

The energies of excited states are
\begin{equation}\label{CFTlev}
\frac{L}{2\pi}E_{\vec n_{1}\vec n_{2}\vec n_{3}\cdots }(L)\simeq -\frac{1 }{4}
+\sum _{k=1}^{N}\sum_{\alpha=1}^{3}| n_{k}^{(\alpha)}|
% +(\sum _{k=1}^{N}N_{k})^{}
% +\frac{1}{4}N^{2}R^{4}(L)
\end{equation}
where
\(\vec n_{k}=(n_k^{(1)},n_k^{(2)},n_k^{(3)})\) are the momentum numbers of particles constituting the state.
We see that the small \(L\) asymptotics of our plots are well described by this formula: The excited states in the   \(U(1)\) sector, denoted in the fig.\ref{pic4} by \(\theta_{n_{1},n_{2}n_{3},\cdots }\), according to the  mode numbers \(n_{1},n_{2},n_{3}\),  approach  the values predicted by (\ref{CFTlev})  (up to the circumstance described in the last footnote). In this sector they   have no excited left and right magnons (no \(u,v\) roots), and only one component is activated:  \(\frac{L}{2\pi}E_{n_{1}n_{2}n_{3}\cdots }(L)=\sum _{k}^{N}n_{k}^{(1)}\).  Say, the curves \(\theta_{0},\theta_{00},\theta_{000},\theta_{0000},\) approach \(-1/4\) at \(L\rightarrow0\), the curves \(\theta_{1},\theta_{01},\theta_{001}\) approach \(3/4\),
the curve \(\theta_{2}\) approaches \(7/4\), etc.
The state with one left and one right magnon excited, denoted as \([u,v,\theta_{-1,1}] \), also approaches \(7/4\).

The qualitative  behavior of the states  \(\theta_{0},\theta_{00},\theta_{000},\theta_{0000}\),etc, at very small \(L\)'s can be explained by the fact that the quantum fields are dominated by their zero modes. \footnote{We would like to thank A.Tsvelik, P.Wiegmann and K.Zarembo for the explanations on this subject.} Since the momentum modes are not excited the field \(\vec Y(\sigma,\tau)\) does not depend on \(\sigma\).
The action and the hamiltonian become:

\begin{equation}
\mathcal{S_{\sigma}}\approx \ \ \frac{L}{e_{0}^{2}(L)} \int dt \,(\partial_{\tau} \phi)^{2},\qquad \hat H=\frac{1}{4} \ \ \frac{e_{0}^{2}(L)}{L}\,\hat J^2
 \end{equation}
  where  the angle \(\phi(\tau) \) represents the coordinate of a material point (a top) on the main circle of the unit sphere, and \(\hat J\) is the corresponding angular momentum.  The quantum mechanical spectrum of this system is well known:
\begin{equation}\label{ZEROMODE}
\frac{L}{2\pi}(E_{\theta_{{\underbrace{\{0,0,\dots,0\}}_{m\,\text{times}} }}}-E_{0})=\frac{1}{8\pi}e_{0}^{2}(L)\,m(m+2)\approx \frac{m(m+2)}{4\log (\omega/ L)}\;
\end{equation}
where $\omega=\gamma+\log\left(\frac{\sqrt e}{\sqrt 8}\frac{\Gamma(3/2)}{4\pi}\right)\simeq
13.66$ is a constant \cite{Shin:1996gi}.
This formula explains well the fact that the corresponding plots on fig.\ref{pic4} converge slowly, as inverse logarithm, to \(-1/4 \) and their spacing is approximately linearly growing with the number \(m\).

The perturbative calculation of the mass gap \([E_{\theta_{0}}(L)-E_{0}(L)]\) for  \(L\ll1\) was done in \cite{Shin:1996gi} and was compared with the numerical results following from the TBA approach in \cite{Balog:2003yr} . Since our numerics is  in a perfect agreement with  \cite{Balog:2003yr}, for the states for which their method works, we will not review it here. We only recall that, in the logarithmic approximation,

\begin{equation}
\frac{L}{2\pi}[E_{\theta_{0}}(L)-E_{0}(L)]\approx\frac{3 }{4}\frac{1}{|\log L|},\qquad(L\ll 1)
\end{equation}which is in the perfect agreement with \eq{ZEROMODE} at m=1.
We also compared \eq{ZEROMODE} for $m=2$ with our numerics and found a good
agreement\footnote{The discrepancy with the r.h.s. of \eq{ZEROMODE} for $m=2$
and $L=1/10,1/100,1/1000$ is $0.057,0.034,0.023$.}.  The inverse logarithm in  \eq{ZEROMODE}  explains well the slight divergence of various curves with zero mode numbers at increasing \(L\)  in fig.\ref{pic4}. Most probably, the divergence of the other plots at increasing \(L\), corresponding to the same value of \(\sum _{k=1}^{N}\sum_{\alpha=1}^{3}| n_{k}^{(\alpha)}|\) can be also perturbatively described by the same inverse log terms.
It would be interesting to study the small $L$ limit analytically to recover
analytic properties of the perturbation theory.

At large \(L\), we enter the realm of the asymptotic Bethe ansatz with L\"uscher-type exponentially small  corrections to these regime. Actually, they describe very well our exact numerical plots for all the considered states, considerably beyond the values of \(L\) allowed by the approximation, as seen in the plot \ref{Luscher:fig}.

\section{ Conclusions }

\begin{figure}[ht]
\epsfxsize=13cm
\centerline{\epsfbox{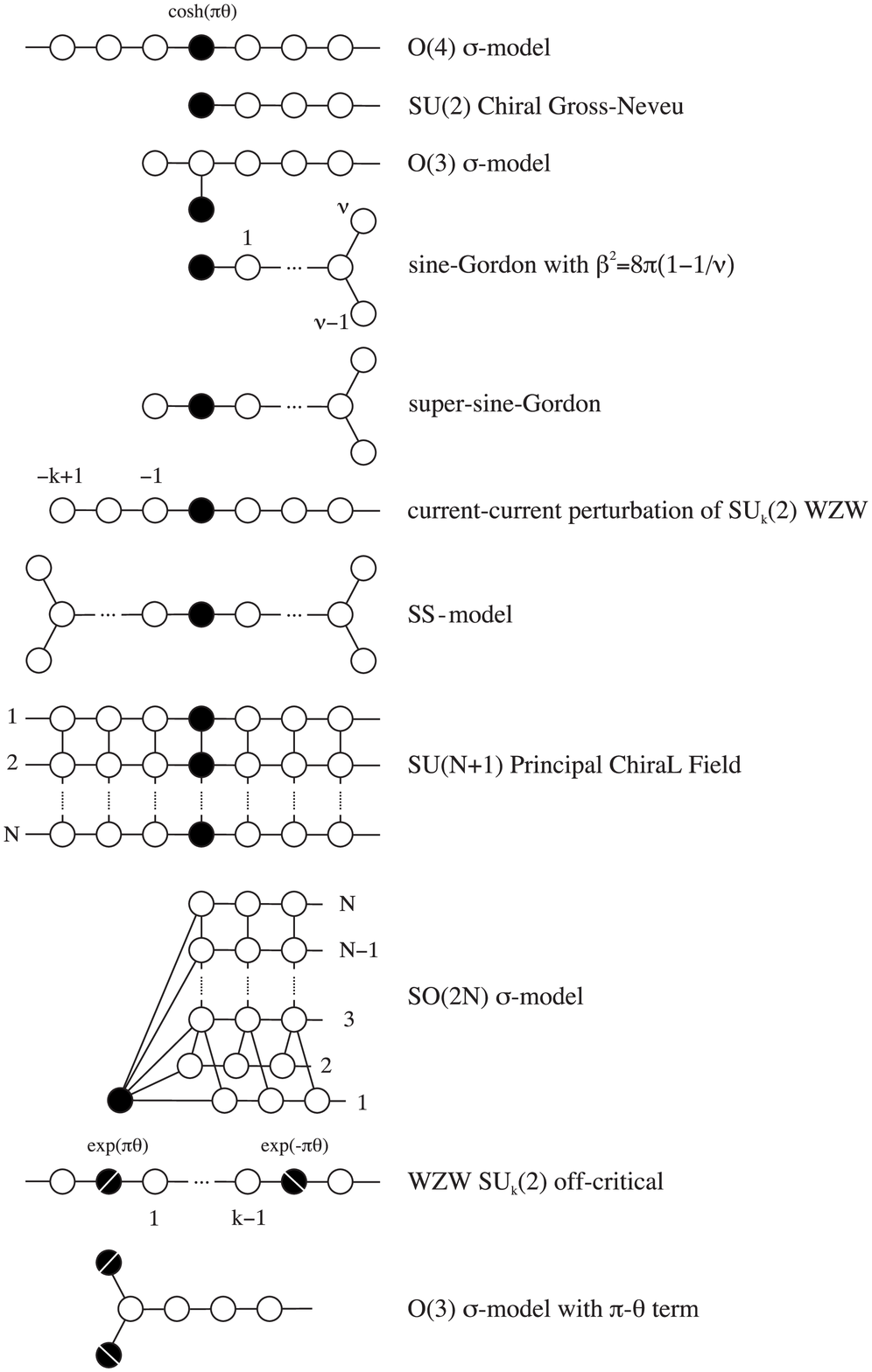}}
\caption{\label{diagrams}
\small\textsl{For several models the Y-system TBA equations for the ground state energy are known and can be represented by the diagrams such as the ones in this figure. Using the techniques developed in this paper it would be extremely interesting to compute their complete spectrum in a systematic way.
} }
\end{figure}

We derived in this paper  the non-linear equations yielding the energy of an arbitrary excited state in the \(O(4)\) \(\) two-dimensional sigma-model, equivalent to the \(SU(2)\) principal chiral field,  defined on  a space circle of an arbitrary  length \(L\) (measured in infinite volume mass gap units). The main formulae we found are reviewed in  subsection \ref{review}.

Although we considered mostly the   \(O(4)\) sigma model the new method which we develop here is very universal and should be applicable to any integrable relativistic sigma model, such as the \(SU(N)_{}\) principle chiral field at any \(N\) , \(O(n)\) sigma models at any \(n\), or more exotic models like SS-model or supersymmetric  Sine-Gordon model (see \cite{Hegedus:2005bg} for the examples and the ground state energy and \cite{Hegedus:2006st} for some excited states). In figure \ref{diagrams} the ground state Y-system diagrams for many known models are represented --  it would be extremely interesting to perform a systematic study of such models using our formalism.

We also hope that the method will  eventually allow to  calculate  the spectrum of finite size operators (such as the Konishi operator) in the N=4 SYM theory, when applied to  its dual, the  integrable string sigma model on AdS\(_{5}\times S^{5}\) background, on a world sheet cylinder of a finite circumference, as inspired by the works \cite{Janik:2006dc,Janik:2007wt,Arutyunov:2007tc,Ambjorn:2005wa}. In spite of the last spectacular applications of the S-matrix approach for the perturbative  calculation of wrapping interactions for Konishi and other twist-2 operators \cite{BJ}, the problem of finding the dimensions of such operators at any coupling is still open.

The Hirota equation, which is equivalent to Y-system,   appears to be a remarkable tool for solving the  integrable
sigma models in 2 space time dimensions. Not only does it help to collapse an infinite system of equations into a few ones, but it also helps to guess the  analytic properties of the remaining unknown  quantities and thus formulate the problem in terms of a single  equation for a complex function  \(g(x)\) \footnote{It might be a few functions for other sigma models but always a finite number of them.
} . This function has the transparent meaning of a gauge transformation between the \(T\) functions in two different solutions to the Hirota equation, but solving the same Y-system.  This equation reminds of the famous Destri-deVega equation, and in the known cases, like the chiral Gross-Neveu model, even coincides with it in certain variables, as we demonstrated in this paper. However, for many interesting   sigma models the Destri-deVega  equation is not known - in particular for  general states in  finite volume systems. Our method suggests a systematic way of deriving  such DdV-like equations.  For example, in the case of \(SU(N)\) symmetry, we can expect that the closed set of such equations should not contain more than 4 real functions (or two complex) - the total number of the gauge functions for the general Hirota equations. It would be interesting to apply our method to the  \(SU(N)\) principal chiral field \cite{Polyakov:1983tt,Wiegmann:PCF,Wiegmann:1984ec}, especially in the large \(N\) limit, which is explicitly solvable for a non-zero magnetic field \cite{Fateev:1994ai}.

There could be interesting applications of our method to  conformal QFT's in two dimensions if we consider them
as some limiting cases of  massive theories.
These limiting cases could be the ultraviolet limit of a small volume \(L\rightarrow \infty\) (see for example \cite{Zhou:1995hg}) or analytic continuation w.r.t. the number of components of a field, like in \cite{Mann:2005ab}, or something else.

Finally, one of the most promising grounds for the applications of our method  should be the case of supersymmetric sigma models, a quickly developing subject, which is very useful in many physical problems ranging from AdS/CFT correspondence to  disordered systems.  The method of solution of Hirota equations applied for the supersymmetric spin chains with the symmetry algebras \(gl(K|M)\), was worked out in \cite{Tsuboi-1,Juttner:1997tc,Kazakov:2007fy,Bazhanov:2008yc}.

\subsection{Summary of the main formulae} \la{review}
In this subsection we summarize our final integral equations in a self-consistent set of formulae. The main example considered in this paper was the   \(O(4)\) sigma model where the particles have  two $SU(2)$ spins as internal degrees of freedom. To compute the exact energy of $N$-particle states with
$J_u$ left spins down (and thus $N-J_u$ left spins up) and $J_v$ right spins
down (and thus $N-J_v$ right spins up) we should solve the single integral equation
on a complex function $g(x)$
\beq
g(x)=i e^{\frac{i}{2}  L \sinh(\pi x)} S(x) \exp\[ s*G_v(x-i/2-i0)-s*G_u(x+i/2+i0) \] \,,  \nn
\eeq
where $*$ stands for convolution, $ s(x)=\frac{1}{2\cosh\pi x}$, $S(x)=\prod_{j=1}^N  S_0(x-\theta_j)$,
$S_0(x)=i\frac{ \Gamma \left(1/2-ix/2\right) \Gamma\left(+ix/2\right)}{\Gamma \left(1/2+ix/2\right) \Gamma \left(-ix/2\right)}$
and the resolvents are given by
\begin{equation}
G_w(x)=\int_{-\infty}^{+\infty}\frac{dy}{2\pi i} \left(
\frac{\rho_w(y)}{x-y-i/2} -
\frac{\bar\rho_w(y)}{x-y+i/2}\right),\quad\;\;w=u,v\,, \nn
\end{equation}
where the densities are parameterized in terms of $g(x)$ as
\beq \nn
e^{\rho_u}=\frac{\frac{g^{+}}{\bar g^+}\Q^{++}_u \Q_v^{++}- \bar \Q_u\bar\Q_v}{\frac{g^{+}}{g^-}\Q_u^{++}\bar \Q_v^{--}-\Q_u\bar\Q_v}
\;\;,\;\;
e^{\rho_v}=\frac{\frac{ g^+}{\bar g^+} \Q_v^{{++}}
   \Q_u^{{++}}-\bar\Q_v \bar \Q_u }{\frac{\bar  g^-}{\bar g^+}\Q_v^{{++}}\bar{\Q}_u^{{--}}
   - \Q_v \bar{\Q}_u  }\;,
\eeq
with $\Q_w(x)=\prod_{k=1}^{J_w}(x-w_k)$ and $\bar\Q_w(x)=\prod_{k=1}^{J_w}(x-\bar w_k)$ being the Baxter functions encoding the Bethe roots of the left and right ``magnons" (\(w=u,v\)). The superscripts $\pm$ indicate the shifts by $\pm i/2$, so that e.g. $\Q_v^{++}=\Q_v(x+i)$ and the bars indicate the complex conjugation. Finally the constants $\theta_j$, $u_j$ and $v_j$ are fixed by the finite volume Bethe equations
\beq
\frac{\Q_u^+(\theta_j)\Q_v^+(\theta_j)}{\bar \Q_u^-(\theta_j) \bar\Q_v^-(\theta_j)}\frac{g(\theta_j)}{\bar g(\theta_j)}=1 \,\, , \,\,
- \frac{\phi^-(u_j) \Q_u^{++}(u_j)}{\phi^+(u_j)\bar \Q_u^{--}(u_j)}P_u(u_j)=1\,\, , \,\,
- \frac{\phi^-(v_j) \Q_v^{++}(v_j)}{\phi^+(v_j)\bar \Q_v^{--}(v_j)}P_v(v_j)=1 \,, \nn
\eeq
where $P_{w=u,v}(x)$ is defined on the upper half plane by
\beq
P_w(x)=\exp \[ K_1(x-i/2)*\rho_w-K_1(x+i/2)*\bar \rho_w\] \,\, , \,\, K_1(x)=\frac{2}{\pi}\frac{1}{4x^2+1} \nn \,,
\eeq
and by its analytic continuation in the full complex plane. The energy of the state is then given by
\beq
E=\sum_{k=1}^{N}\cosh(\pi\theta_{k})\,-\frac{1}{2}\,\int_{-\infty}^{+\infty} \cosh(\pi x) (\rho_u(x)+\bar\rho_u(x)) dx\;. \nn
\eeq

\section*{Acknowledgements}

The work of NG was partly supported by the German Science Foundation (DFG) under
the Collaborative Research Center (SFB) 676.  The work of NG and VK was partly supported by  the ANR grant INT-AdS/CFT
(contract ANR36ADSCSTZ) and the grant RFFI 08-02-00287. The work of V.K. is also partly supported by the ANR grant GranMA BLAN-08-1-313695. N.G. and P.V. would like to thank  Ecole Normale Superieure where a substantial part of this work was done. N.G. would also like to thank Commissariat ˆ l'Energie Atomique where a substantial part of this work was done. V.K. would like to thank  I.Kostov and  Z.Tsuboi, P.Wiegmann for the useful discussions, and especially Anton Zabrodin, who shared with him his profound knowledge of discrete classical integrability. He also thanks the AEI (Potsdam) for the kind hospitality  during the work on important parts of the project and the Humboldt fundation for the support.  We also would like to thank A.Tsvelik,  J.Teschner, V.Schomerus, K.Zarembo for many enlightening discussions, as well as A.Kozak who participated in  this project on its the early stage.

%%%%%%%%%%%%%%%%%%%%%%%%%%%%%%%
%%%%%%%%%%%%%%%%%%%%%%%%%%%%%%%
%%%%%%%%%%%%%%%%%%%%%%%%%%%%%%%
%%%%%%%%%%%%%%%%%%%%%%%%%%%%%%%
%%%%%%%%%%%%%%%%%%%%%%%%%%%%%%%
%%%%%%%%%%%%%%%%%%%%%%%%%%%%%%%
%%%%%%%%%%%%%%%%%%%%%%%%%%%%%%%
%%%%%%%%%%%%%%%%%%%%%%%%%%%%%%%
%%%%%%%%%%%%%%%%%%%%%%%%%%%%%%%1

\appendix

\section{Derivation of the (ground state) $Y$-system}\label{AppA}

To compute the ground state energy $E_0(L)$ of the $SU(2)$ principal
chiral field in a periodic box of a size L we can compute its Euclidean path integral $Z$ with the
fields living on the space-time torus of the size \(L\times R\) , where the periodic imaginary  ``time"  $R  $
 is very big\beq
Z=e^{-R E_0(L)} \,.
\eeq
Following Al.Zamolodchikov  \cite{Zamolodchikov:TBA1990} we can compute
this quantity   exchanging the
role of $L$ and $R$ so that the latter becomes the space variable
whereas the former becomes the periodic time. Since $R\to \infty$
the spectrum corresponding to the new Hamiltonian can be computed
from the asymptotic Bethe ansatz and the finite periodic time $L$
means that we should consider the system at a finite inverse temperature $L$.
Thus we conclude that
\beq
E_0(L)=f(L) \,.
\eeq
where $f(L)$ is the free energy per unit length of the $SU(2)$ PCF
at the  temperature $1/L$ in the thermodynamical limit, when $R\to
\infty$.

To compute the free energy we will start by reviewing the asymptotic
spectrum of the theory as given by the asymptotic Bethe ansatz. Then
we will recall what are the magnon bound states (complexes, or strings) and how they
are organized in the complex plane. We will see that the quantum
states in the thermodynamic limit can be described by the
densities of these  complexes and their holes.  From this
description we will  write  the entropy formula and
 thus find the desired exact free energy as the result of the saddle point approximation at  $R\to
\infty$. This will give the TBA equations.

Particles in the $SU(2)$ principal chiral field transform in the
bi-fundamental representation under two $SU(2)$ groups. The theory is integrable and
thus the general scattering process factorizes into a sequence of
two-body scattering events. The $S$-matrix \cite{Zamolodchikov:1978xm} describing the scattering
of two particles with momenta and energies given by
\beq
p_j=m\sinh (\pi \theta_j)\;\;,\;\; E_j=m \cosh(\pi\theta_j) \,\, ,
\eeq
depends only on the difference of rapidities
$\theta=\theta_1-\theta_2$
\beq
\hat {S}_{12}(\theta) =S_0(\theta) \frac{\hat R(\theta)}{\theta-i} \otimes S_0(\theta)  \frac{\hat R(\theta)}{\theta-i} \,\, , \, \, S_0(\theta)=i\frac{ \Gamma \left(\frac{1}{2}-\frac{i \theta}{2}\right) \Gamma
   \left(+\frac{i \theta}{2}\right)}{\Gamma \left(\frac{1}{2}+\frac{i
   \theta}{2}\right) \Gamma \left(-\frac{i \theta}{2}\right)}\;,
\eeq
where $\hat R(\theta)$ is the usual $SU(2)$ R-matrix in the
fundamental representation given by
\beq
\hat R(\theta)=\theta+i P\;,
\eeq
where $P$ is the permutation operator exchanging the spins of the
scattered particles.

From now on, we will measure the length \(L\) in the units of the mass gap \(m\), which means that we will put \(m=1\).

When $N$ particles are put on a large circle of length $R$ the
periodicity condition to be imposed on the wave function  reads
\beq
-\hat{\mathcal{T}} (\theta_j)\,e^{iR\sinh(\pi\theta_j)} \Psi=\Psi\;,  \la{Psi}
\eeq
where $\mathcal{T}$ is the  transfer matrix
\beq
\hat{\mathcal{T}} (\theta)\equiv {\rm tr}_0 \left(\hat{S}_{01}\left(\theta-\theta_1\right)\dots \hat{S}_{0N}\left(\theta-\theta_N)\right) \right)\;,
\eeq
with the index $0$ for an additional auxiliary particle which we
scatter against all physical particles. The trace is taken over this
auxiliary space. Indeed, when the transfer matrix is evaluated at a
value of the physical rapidity $\theta_j$  the
corresponding $S$-matrix $\hat S_{0j}(\theta-\theta_j)$ becomes
simply $-P_{0j}\otimes P_{0j}$ which means that the auxiliary
particle changes the quantum numbers and becomes the physical
particle $\theta_j$. Then (\ref{Psi}) becomes the periodicity
condition (\ref{perintro}) which physically states that once we pick
the particle $j$ and carry it around the circle the total phase acquired
by the wave function -- which will be given by the free propagation
$R p_j$ plus the phase shifts do to the (factorized) scattering with
each of the other particles -- must be a trivial multiple of $2\pi$.

Using the algebraic or analytic  Bethe ansatz technologies it is possible to
diagonalize $\mathcal{T}(\theta)$ for any value of $\theta$ using
the same eigenvector basis (see an Appendix from \cite{Gromov:2006dh} for the details). Multi-particle states with $J_u$ left
spins down (and thus $N-J_u$ left spins up) and $J_v$ right spins
down (and thus $N-J_v$ right spins up) are parameterized by $J_u$
auxiliary Bethe roots $u_j$ and $J_v$ roots $v_j$ and
\beq
\hat{\mathcal{T}}(\theta)\Psi=\frac{S^2(\theta)}{\phi^2(\theta-i)} T_1^u(\theta-i/2)T_1^v(\theta-i/2) \Psi\;,
\eeq
% with
% \beq
% \Lambda(\theta)=\frac{S^2(\theta)}{\phi^2(\theta-i)} T_1^u(\theta-i/2)T_1^v(\theta-i/2) \,,
% \eeq
where $T_1^{u(v)}$ is the transfer matrix in the fundamental
representation associated with the left (right) $SU(2)$ spins,
\beq
T_1^w(\theta)\equiv
\frac{\Q_w(\theta+i)\phi(\theta-i/2)+\Q_w(\theta-i)\phi(\theta+i/2)}{\Q_w(\theta)}
\,, \la{T1}
\eeq
and
\beq
\phi(\theta)\equiv \prod_{j=1}^N (\theta-\theta_j)\,\,\, ,\,\,\, S(\theta)=\prod_{j=1}^N S_0(\theta-\theta_j)\\,\,\,\,\,\, \Q_u(\theta)=\prod_{j=1}^{J_u}(\theta-u_j),\,\,\,\,\, \Q_v(\theta)=\prod_{j=1}^{J_v}(\theta-v_j)\,.
\eeq
The  rapidities $\theta_j$ and $u_j,v_j$ are then fixed by a
set of nested Bethe equations. The Bethe equations for the physical
rapidities $\theta_j$ are given by the periodicity condition
(\ref{Psi}) which can be written as
\beqa
e^{-imR\sinh(\pi\theta_j)}=-\frac{S^2(\theta_j)}{\phi^2(\theta_j-i)}
T_1^u(\theta_j-i/2)T_1^v(\theta_j-i/2)\;,
\eeqa
or simply
\beqa
e^{-imR\sinh(\pi\theta_j)} = -S^2(\theta_j)
\frac{\Q_u(\theta_j+i/2)}{\Q_u(\theta_j-i/2)}
\frac{\Q_v(\theta_j+i/2)}{\Q_v(\theta_j-i/2)} \,. \label{BAEmiddle}
\eeqa
The magnon rapidities $u_j$ and $v_j$ are fixed by the auxiliary
Bethe equations
\beqa
-\frac{\Q_u(u_j+i)}{\Q_u(u_j-i)} =\frac{\phi(u_j+i/2)}{\phi(u_j-i/2)}   \,\, , \,\, -\frac{\Q_v(v_j+i)}{\Q_v(v_j-i)} =\frac{\phi(v_j+i/2)}{\phi(v_j-i/2)}\;, \la{BAEuv}
\eeqa
which appear in the diagonalization of the left and right transfer
matrices. Notice that these equations ensure that the apparent poles
in (\ref{T1}) drop out and render $T_1^w(\theta)$ polynomial as it
ought to be. For each solution to these equations we obtain the
energy of the corresponding quantum state from
\beqa
E=\sum_{j=1}^N  \cosh(\pi \theta_j)\,.
\eeqa

To be able to compute the free energy $f(L)$ we need to understand
how the solutions to these Bethe equations organize themselves so
that we  can introduce the entropy density. Let us consider the
auxiliary roots $u$, obviously the same considerations will apply for the
$v$ roots. These roots can take complex values. When $u_j$ has a
positive imaginary part the r.h.s of the Bethe equations in
(\ref{BAEuv}) diverges,
\beq
\frac{\phi(u_j+i/2)}{\phi(u_j-i/2)} \stackrel{N\to \infty}{\to} \infty\;,
\eeq
which means that
\beqa
\frac{\Q_u(u_j+i)}{\Q_u(u_j-i)}
\eeqa
must diverge as well. This can be achieved if there is another
magnon rapidity $u_k$ such that $u_j-u_k\simeq i$. Thus, in the
thermodynamical limit the magnon rapidities will organize themselves
into a Bethe-string of $n$ roots $u_j$ spaced by $i$. In particular,
a single real root corresponds to a Bethe string with $n=1$. The
Bethe equations can then be multiplied for $u_j$'s belonging to the
same string so that this gives  new  Bethe equations,
solely for the (real) center of each string. This is the usual fusion
procedure applied at the level of the Bethe equations. The resulting
equations look as follows. Introduce the magnon bound states:
\begin{eqnarray*}
u_{j,a}^{(n)}=u_{j}^{(n)}+i\frac{1}{2}(n+1)-ia,\quad a=1,\dots,n.
\end{eqnarray*}
Multiplying the equations for a given $n$-bound state we get for
\eq{BAEuv} and
\eq{BAEmiddle}
\begin{eqnarray*}
e^{-iR p(\theta_\alpha) }&=& \prod_{\beta\neq \alpha}\, S_0^{\,2}
\left( \th_\a-\th_\b \right) \prod_{j,n}
\frac{\th_\a-u_{j}^{(n)}+i\frac{n}{2}}{\th_\a-u_{j}^{(n)}-i\frac{n}{2}}\,,\\
\prod_\b\frac{u_{j}^{(n)}-\th_\b+i\frac{n}{2}}{u_{j}^{(n)}-\th_\b-i\frac{n}{2}} &=&\!\!\!\!
\prod_{(k,m)\neq(j,n)}
\frac{u_{j}^{(n)}-u_{k}^{(m)}-i\frac{n+m}{2}}{u_{j}^{(n)}-u_{k}^{(m)}+i\frac{n+m}{2}}\cdot
\frac{u_{j}^{(n)}-u_{k}^{(m)}-i\frac{|n-m|}{2}}{u_{j}^{(n)}-u_{k}^{(m)}
+i\frac{|n-m|}{2}} \prod_{s=\frac{|n-m|}{2}}^{\frac{n+m}{2}}
\left(\frac{u_{j}^{(n)}-u_{k}^{(m)}+is}{u_{j}^{(n)}-u_{k}^{(m)}-is}\right)^2\,.
\end{eqnarray*}
In the thermodynamic limit we will have a large number of each type
of Bethe roots which we can describe by a density $\varrho_n$. We use
$n=0$ for the density of $\theta$ particles, $n \ge 1$ to describe
the density of $u$ Bethe strings of size $n$ and $n\le -1$ for the
$v$ Bethe strings made out of $-n$ roots. For each density of
particles we also have the corresponding density of holes $\bar
\varrho_n$. Bethe equations in the thermodynamic limit, obtained by taking the logarithmic derivatives of both sides of these equations,  read
\beq
\varrho_n+\bar\varrho_n=\frac{R}{2} \cosh(\pi\theta) \delta_{n0} -\sum_{m=-\infty}^{\infty} K_{n,m}*\varrho_m\;,
\eeq
where $*$ stands for the usual convolution
\beq
f*g=\int_{-\infty}^{+\infty} d\theta' f(\theta-\theta')g(\theta')\;,
\eeq
and $K_{nm}$ is the derivative of the logarithm of the effective
$S$-matrix between the strings of size $n$ and $m$. In particular we have
\beq\label{eq:K0}
-K_{0,0}(\theta)\equiv K_0(\theta)=\frac{1}{2\pi i} \frac{d}{d\theta}
\log S_0^2(\theta)\;,
\eeq
for the interaction between physical rapidities,
\beq
K_{0,n}(\theta)=-K_{n,0}(\theta)=\frac{1}{2\pi i} \frac{d}{d\theta}
\log \frac{\theta-i|n|/2}{\theta+i|n|/2} =\frac{1}{\pi}
\frac{2|n|}{4\theta^2+|n|^2}\equiv K_n(\theta)\;\;,\;\; n\neq0\;,
\eeq
for the interaction of $\theta$ rapidities with Bethe strings of
size $|n|$ and
\beq
K_{n,m}(\theta)
=K_{-n,-m}(\theta)=\sum\limits_{s=\frac{|n-m|}{2}+1}^{\frac{n+m}{2}}2K_{2s}(\theta)
-K_{n+m}(\theta) +K_{|n-m|}(\theta) \delta_{n\neq m} \,\,, \,\,
n,m=1,2,\dots\;, \la{Knmspace}
\eeq
for the interaction of two Bethe strings. Obviously $K_{n,m}=0$ if
$n\times m < 0$.

It is interesting that even though these kernels appear as some
quite complicated functions they all exhibit very simple fourier
transforms $\hat K_{n,m}$. More precisely we have
\beq
\hat K_0(\omega)
=\frac{e^{-|\omega|/2} }{\cosh\frac{\omega}{2}} \,,
\eeq
and
\beq
\hat K_n(\omega)
=e^{-|n| \omega/2} \,\, , \,\, n=1,2,\dots\;,\,
\eeq
so that the sum in (\ref{Knmspace}) can be explicitly done yielding
\beq
\hat K_{n,m}=\coth\left(\frac{|\omega|}{2}\right) \left(e^{-\frac{|\omega|}{2} |m-n|}-e^{-\frac{|\omega|}{2}(m+n)}\right)-\delta_{n,m} \,\, , \,\, n,m=1,2,\dots\,
\eeq

A  very useful formula for what follows concerns the
inversion of the operator $K_{nm}$ when both indices $n$ and $m$ are
restricted to be positive (or negative). In Fourier space
\beq
(\hat K_{nm}+\delta_{nm})^{-1} =
\delta_{mn}-\hat{s}\left(\delta_{n,m+1}+\delta_{n,m-1} \right), \,\,\quad(n,m>0)\;, \label{Km1}
\eeq
where the operator $\hat{s}$  (and its  fourier transform)
has the following form\beq
s(\theta)=\frac{1}{2\cosh \pi \theta} \,,\quad\,\,
\left(\hat{s}(\omega)=\frac{1}{2\cosh\left(\frac{\omega}{2}\right)}\right).
\eeq
In particular we notice that
\begin{equation}\label{K0K1}
 K_0=2s*K_1.
\end{equation}

Having introduced all the necessary kernels we can proceed to
construct the quantity of interest, the free energy at
the temperature $1/L$. We have
\beq
f(L)={\rm min}_{\varrho_n,\bar \varrho_n} \int d\theta\left( \varrho_0   L \cosh
\pi\theta- \sum_{n=-\infty}^{\infty} \varrho_n\log\left(1+\frac{\bar
\varrho_n}{\varrho_n}\right)+ \bar\varrho_n\log\left(1+\frac{ \varrho_n}{\bar\varrho_n}\right)\right)\;,
\eeq
where we should minimize the integral by varying the densities of
particles and holes keeping the Bethe equations satisfied. The first
term in the integral is the energy density multiplied by the inverse
temperature (which is $L$) and the second term given by the sum over
$n$ is the entropy density (see e.g. \cite{Zamolodchikov:TBA1990}). We use Bethe
equations to write the variation $\delta
\bar\varrho_n=-\delta\varrho_n-\sum_{m=-\infty}^{\infty} K_{nm}*\delta
\varrho_m$ so that the extremum condition $\delta f=0$ yields a set of
TBA equations
\begin{eqnarray}\label{FTBA}
0=-\epsilon_n+ L \cosh(\pi\theta) \delta_{n,0}+  \sum_{m=-\infty}^{\infty} K_{mn} * \log\left(1+e^{-\epsilon_m}\right)\;,  \la{pre}
\end{eqnarray}
where $\frac{\varrho_n}{\bar\varrho_n}=e^{-\epsilon_n}$ . The free energy
evaluated at this extremum can then be written in terms of
$\epsilon_0$ alone, so that
\begin{equation}
E_0(L)= -   \int \frac{d\theta}{2}  \cosh(\pi\theta)  \log\left(1+e^{-\epsilon_0(\theta)}\right)\;.
\end{equation}
The last two equations yield the finite size ground state energy of the
$SU(2)$ principal chiral field.

Finally, we will show below that defining the incidence matrix $I_{nm}=\delta_{n,m\pm 1}$,
the $Y$ variables $ Y_m=e^{\epsilon_m} \,\, (m\neq0), \,\,
Y_0=e^{-\epsilon_0} $ and using the  operator (\ref{Km1})
these equations can be transformed  into a local (in the discrete variable $n$) set of integral equations
\begin{equation}\label{YTBA}
\log\left(Y_n\right)+ L  \cosh(\pi\theta)\,\delta_{n,0}=\sum_{m=-\infty}^\infty I_{nm} \,s*\log\left(1+Y_m\right) \,,\qquad-\infty<n<\infty
\end{equation}
and
\begin{equation}
E_0(L)= - \int \frac{d\theta}{2}   \cosh(\pi\theta)  \log\left(1+Y_0 \right) \,.
\end{equation}
It is remarkable that  non-local kernels $K_{nm}$ disappear and
at the end  only a very simple kernel \( \hat s \) appears in the
final set of TBA equations. It is also remarkable, and still somewhat mysterious, that the \(SU(2)_{L  }\) and \(SU(2)_{R}\) wings are smoothly glued into  one Y-system on a discrete set \(-\infty<n<\infty\).\

To show \eqref{YTBA} we should consider \eqref{FTBA} separately for \(n>0\), \(n<0\) and \(n=0\). Applying the operator  (\ref{Km1})
to  \eqref{FTBA} in the first two cases (it is convenient to rearrange them to have the combination \(\delta _{mn}+K_{mn}\)), we easily verify \eqref{YTBA}, except the case \(n=0\) which we should consider separately. To find  $n=0$ equation of the Y-system we consider (\ref{FTBA}) for $n=-1,0,1$. The kernels $K_{n,m}$ entering these three equations are
\beq
-K_{0,0}=K_0=2s*K_1\,\, , \,\, K_{0,\pm 1}=K_1 \,\, ,\,\,  K_{\pm m,0}=-K_{m} \,\, , \,\, m>0
\eeq
and, most importantly,
\beqa
K_{\pm m, \pm 1}=K_{m+1}+ K_{m-1} \delta_{m\neq 1} \,. \la{kronecker}
\eeqa
Thus if we convolute (\ref{FTBA}) for $n=1$ with the inverse shift operator $s$ and use that $s*(K_{m+1}+K_{m-1})=K_m$ we get
\begin{eqnarray}
%0= -s*\epsilon_{1} -  s*K_1*\log\left(1+e^{-\epsilon_0}\right)  +\sum_{m=1}^{\infty} s *\left(K_{m+1}+ K_{m-1}- K_{m-1} \delta_{m,1}\right)* \log\left(1+e^{-\epsilon_m}\right)  \\
0=  -s*K_1*\log\left(1+e^{-\epsilon_0}\right)  +\sum_{m=1}^{\infty}  K_{m}*\log\left(1+e^{-\epsilon_m}\right)- s*  \log\left(1+e^{+\epsilon_1}\right) \la{equ1}  \,.  \end{eqnarray}
Notice that the last term is separated from the infinite sum because the $m=1$ case in (\ref{kronecker}) behaves slightly differently than for the other $m$'s. Moreover the sign of the exponent inside this log differs from that inside the logs in the infinite sum because we absorbed the first term in (\ref{FTBA}) into this last log. Similarly, for $n=-1$ we have
\begin{eqnarray}
%0= -s*\epsilon_{1} +  s*K_1*\log\left(1+e^{-\epsilon_0}\right)  +\sum_{m=1}^{\infty} s *\left(K_{m+1}+ K_{m-1}- K_{m-1} \delta_{m,1}\right)* \log\left(1+e^{-\epsilon_m}\right)  \\
0=  -s*K_1*\log\left(1+e^{-\epsilon_0}\right)  +\sum_{m=-\infty}^{-1}  K_{|m|}*\log\left(1+e^{-\epsilon_m}\right)- s*  \log\left(1+e^{+\epsilon_{-1}}\right)   \,. \la{equ2}
\end{eqnarray}
These two equations can then be used to simplify the $n=0$ equation which reads
\begin{eqnarray}
0= \epsilon_{0} - L \cosh(\pi\theta)+ K_0*\log\left(1+e^{-\epsilon_0}\right)  -\sum_{m\neq 0} K_{|m|}* \log\left(1+e^{-\epsilon_m}\right) \,. \la{equ3}
\end{eqnarray}
Indeed if we sum all these three equations we see that (i) the infinite sums completely cancel out, (ii) the convolutions with $\log\left(1+e^{-\epsilon_0}\right)$ drop out as well by virtue of the  identity $K_0=2s*K_1$. We are thus left with the last terms in (\ref{equ1}) and (\ref{equ2}) plus the first two terms in (\ref{equ3}) thus obtaining the last $Y$-system equation (\ref{YTBA}) for $n=0$.

Moreover, for the functions $g(\theta)$ analytic inside the physical
strip $\IM(\theta)<1/2$ we have
\begin{equation}
s*\[g(\theta+i/2)+g(\theta-i/2)\]=g(\theta)
\end{equation}
because
\begin{eqnarray}
&&\nn\int_{R}\frac{g(\theta+i/2)+g(\theta-i/2)}{2\cosh(\pi(\theta-x))}d\theta=
%\int_{R+i/2}\frac{g(\theta)}{-2i\sinh(\pi(\theta-x))}d\theta+
%\int_{R-i/2}\frac{g(\theta)}{+2i\sinh(\pi(\theta-x))}d\theta\\
%&&=
\frac{1}{2i}\ccw\oint_{\rm PS} \frac{g(\theta)}{\sinh(\pi(\theta-x))}d\theta=g(x)\,.
\end{eqnarray}
Therefore if $Y_n$ is non-zero inside the physical strip we can
invert the $s$ operator to find a set of functional equations, finally rendering the  \(Y\)-system for the PCF at a finite temperature \(1/L\)
\begin{equation}
Y_n(\theta+i/2)Y_n(\theta-i/2)=(1+Y_{n-1}(\theta)) (1+Y_{n+1}(\theta))
\end{equation}
To fix a solution,  this \(Y\)-system   ought to be supplemented by the large $\theta$ asymptotics
$Y_{n}\simeq e^{-\delta_{n0}\,L\cosh\pi\theta}$ related to the relativistic dispersion relation. Notice that these
functional equations do not contain the dispersion relation explicitly:  it
appears only through the asymptotics of the $Y$-functions.

\section{General solution in terms of Hirota functions}\label{GenSol2}

In this appendix, we will give  an alternative approach to the construction of solution for the energy   of a general  state of the \(SU(2)\) PCF in the periodic box. It will give a new NLIE defining the spectrum, different from  the one of the section \ref{GEN}. This approach is the generalization of the approach we used in section \ref{subsec:U1} for the states with $N$ particles and non-trivial polarizations encoded in the Baxter polynomials $\Q_u$ and $\Q_v$.

As explained in the beginning of the section \ref{GEN}, for each solution to the Y-system equations, there are two natural solutions to Hirota equation which yield the same $Y$'s and are related by a gauge transformation. The expected analytic properties of these functions are described in this section. In particular we have
\begin{eqnarray}\label{eq:GaTTF}
T_{+1}^v&=&g^+\bar g^- T_{+1}^u\nn\\
T_{-1}^u&=&\frac{1}{g^-\bar g^+} T_{-1}^v\nn\\
\Phi_v&=&g^+ g^- \Phi_u
\end{eqnarray}
which we will now use to completely solve our problem. First, as in the section \ref{GEN}, we find the gauge function $g$ from the last relation,
\begin{equation}
g=ie^{iL\sinh(\pi x)/2}\left(\frac{\Phi^v}{\Phi^u}\right)^{*s},\quad\;\;\bar
g=-ie^{-iL\sinh(\pi x)/2}\left(\frac{\bPhi^v}{\bPhi^u}\right)^{*s}
.\end{equation}
and plug it in the first two to find
\begin{eqnarray}\la{Tgen}
T_{+1}^v &=& %e^{-L\cosh(\pi x)}\left(\frac{\Phi_v^+\bar\Phi_v^-}{\Phi_u^+\bar\Phi_u^-}\right)^{*s}T_{+1}^u=
e^{-L\cosh(\pi x)}
 \frac{\Phi_v\bar\Phi_v T_{+1}^u}{(\Phi_u^+\Phi_v^-\bar\Phi_v^+\bar\Phi_u^-)^{*s}}\nn \,,\\
T_{-1}^u &=& e^{-L\cosh(\pi x)} \frac{\Phi_u\bar\Phi_u
T_{-1}^v}{(\Phi_u^+\Phi_v^-\bar\Phi_v^+\bar\Phi_u^-)^{*s}} \,.
\end{eqnarray}
As in the section \ref{GEN}, we still have to relate $T_{0}^{u,v}$ and
$\Phi^{u,v}$, but we do it here by a different relation.  For that, let us define in the whole complex plane
$x$ the functions
\begin{equation}\la{eq:intFu}
F_u(x)=\phi(x)-\frac{\phi(x)}{2\pi i}\int \frac{T^u_{-1}(y)\Q_u(y+i)/\Q_u(y)}{(y-x+i/2)\phi(y+i/2)}dy
+\frac{\phi(x)}{2\pi i}\int \frac{T^u_{-1}(y)\bar \Q_u(y-i)/\bar \Q_u(y)}{(y-x-i/2)\phi(y-i/2)}dy
\end{equation}
and
\begin{equation}\la{eq:intFv}
F_v(x)=\phi(x)-\frac{\phi(x)}{2\pi i}\int
\frac{T^v_{+1}(y) \Q_v(y+i)/\Q_v(y)}{(y-x+i/2)\phi(y+i/2) }dy
+\frac{\phi(x)}{2\pi i}\int
 \frac{T^v_{+1}(y)\bar\Q_v(y-i)/\bar \Q_v(y)}{(y-x-i/2)\phi(y-i/2)}dy\;,
\end{equation}
where the integrals essentially go along the real axis, but we
should pass the contour in such a way that   the zeroes  $u_j,v_j$  of
$\Q_{u,v }$ remain below the contour  and the complex conjugated
zeroes   $\bar u_j$  of
 $\bar \Q_{u,v}$ stay above the contour. Using
\eq{JUMPS} we can show that $T_{0}^{u,v}$ and  $\Phi_{u,v}$ are related to the
values of the same analytic function $F_{u,v}$ inside and outside
of the analyticity strip, respectively:
\begin{equation}\la{eq:Fu}
F_u(x)=\left\{
\bea{ll}
 \frac{\bar \Q_u^-}{\Q_u^-}\Phi^-_u & {\rm Im} x> +1/2\\ \\
T^u_0(x)& |{\rm Im} x|< 1/2\\ \\
\frac{\Q_u^+}{\bar \Q_u^+}\bPhi^+_u & {\rm Im} x< -1/2
\eea
\right.
\,\, \,,\, \,\, F_v(x)=\left\{
\bea{ll}
 -\frac{ \bar \Q_v^-}{\Q_v^{-}}\bar\Phi_v^- & {\rm Im} x> +1/2\\ \\
\quad T^v_0(x)& |{\rm Im} x|< 1/2\\ \\
-\frac{\Q_v^+}{ \bar \Q_v^{+}}\Phi_v^+ & {\rm Im} x< -1/2
\eea
\right.\;.
\end{equation}
 Indeed, substituting  from \eqref{JUMPS} $T^{u}_{-1} =(T_0^{u+} \Q^{u}-\Phi^{u}\bar \Q^{u})/\Q^{u++}$  and the conjugate  $T_{-1}^{u}=(T_{0}^{u-}\bar \Q^{u}-\bar \Phi^{u} \Q^{u})/\bar Q^{u--}$   into the first and second terms of \eqref{eq:intFu}, respectively,  we can separate the contribution of two integrals in \eqref{eq:intFu} containing $T_0^u$  into a single contour integral going around the physical strip (to realize it is useful to make shifts  by $\pm i/2$ for the integration variable). We shall also use the fact that   \(\frac{T_0^u}{\phi^+}\to 1\)   at $x\rightarrow\pm\infty$. If the point $x$ is inside the physical strip we can contract the contour around the pole at $y=x$ and thus verify the middle relation in \eq{eq:Fu}. The other two integrals in \eqref{eq:Fu}, containing   $\Phi$ and
$\bPhi$,  do not contribute since we can close the contours there
around the upper and lower half-plane, respectively. The result is
zero since there are no singularities inside by our previous
assumption. The poles related to zeroes of \( \Q_{u,v}\) do not contribute since
they are outside of these contours by  definition.      We
should ensure by hand the analyticity of $\Phi(x)$ close to the
real axis
\begin{equation}\la{eq:BAEu}
T_0^{u+}\Q^u=T_{-1}^u \Q^{u++}\;\;,\;\;x=\bar u_j\;,
\end{equation}
which is the finite $L$ deformation of the usual asymptotic axillary
BAE for the $u$-roots. The relations for $F_v$ are found in a similar way.
This equation can be shown to be equivalent to the \eq{baeu0} derived in the main
text.

 Notice that $T_0^u$ and $T_0^v$ are automatically
analytic even slightly outside of  the physical strip, because we
can deform the contours to open further the physical strip. We will discuss
this ``extra" analyticity in the App.\ref{ANALYT}.

Finally, notice also that
\beqa
T_1^u(x)&=&\frac{\Q_u(x+i)}{\Q_u(x)}F_u(x-i/2)+\frac{\bar\Q_u(x-i)}{\bar \Q_u(x)}F_u(x+i/2) \,, \nn\\
T_{-1}^v(x)&=&\frac{\Q_v(x+i)}{\Q_v(x)}F_v(x-i/2)+\frac{\bar\Q_v(x-i)}{\bar \Q_v(x)}F_v(x+i/2)\;, \nn
\eeqa
so that (\ref{Tgen}) completely constrain the functions $T_{-1}^u$ and $T_1^v$ out of which all other $T_k$ and $Y_k$ can be written down using the resolvents $F_u$ and $F_v$.

In the $U(1)$ sector we have $\Q_u=\Q_v=1$ and the two wings are obviously equivalent. We will have in this case $T_{1}^v=T_{-1}^u$ and $F_u=F_v$ will be given by (\ref{ansatz}). Equations (\ref{Tgen}) then reduce to the previously derived equation (\ref{eqF}).

\subsection{The main Bethe equations}
The main BAE reflecting the periodicity of the wave function and
constraining the real zeroes $\theta_{j}$  is given by
\eqref{Y0m1:eq}:
\begin{equation}
Y_0(\theta_j\pm i/2)=\frac{T^u_{1}(\theta_j\pm i/2)T^u_{-1}(\theta_j\pm i/2)}{\Phi_u\left(\theta_j\pm i/2\frac{}{}\right)\bPhi_u\left(\theta_j\pm i/2\right)}
=-1\;.
\end{equation}
Using the fact that $T^u_{1}$ has no zeros inside the physical strip
and the denominator is regular for real $\theta_{j}$ we conclude
that $
T^{u}_{-1}(\theta_j\pm i/2)\neq 0
$. This condition can be in fact interpreted as yet another form of the
main BAE. It can be further simplified: using \eqref{eq:GaTTF} we
get
\begin{eqnarray}\la{TmL}
T_{-1}^{u+}&=&\frac{1}{g\bar g^{++}}T_{-1}^{v+}=-
\frac{\bar g\,\bar \Phi^{+}_u}{g\,\bar\Phi^{+}_v}T_{-1}^{v+}\;,%\\
%&\simeq& - e^{-i L\sinh(\pi x)}
%\frac{\bar \Phi^{+}_u}{\bar\Phi^{+}_v}T_{-1}^v(x+i/2)\\
%T_{+1}^v(x+i/2)
%&\simeq& - e^{-i L\sinh(\pi x)}
%\frac{ \Phi^{+}_v}{\Phi^{+}_u}T_{+1}^u(x+i/2)
\end{eqnarray}
where we can use that, due to \eqref{JUMPS}, for the $u$-wing
\begin{eqnarray}
\bPhi_u^+&=&+T^u_0\frac{\bar \Q_u^+}{\Q_u^+}- T_{-1}^{u+}\frac{\bar \Q^{-}_u}{\Q_u^+}\;.%\\
%\Phi_v^+&=& -T^v_0\frac{Q^{+++}_v}{\bar Q_v^{-}}+T_1^{v+}\frac{Q_v^{+}}{\bar Q^{-}_v}\;.
\eeqa
Substituting it into the \eq{TmL}, and evaluating at $\theta_{j}$ which is a zero of $T_0$ we
get
\begin{equation}
1=\frac{\bar g}{g }\frac{\bar \Q_u^-}{\Q_u^+}\frac{T_{-1}^{v+}}{\bar
\Phi_v^+}\vert_{x=\th_j}\;,
\end{equation}
or, using   \eqref{JUMPS} for the $v$-wing,
 we get the simplest form of the main equation, easy to compare with the large $L$ limit
\begin{equation}
1=\frac{\bar g}{g }\frac{\bar \Q_u^-}{\Q_u^+}\frac{ \bar\Q_v^-}{
\Q_v^+}\vert_{x=\th_j}\;.
\end{equation}
which was also derived in the main text \eq{baeth}.

\section{Proof of reality of $T_k$}

In this appendix we shall analyze the reality of the $T$-functions. This is an important point to consider because the Hirota equation is solved explicitly by \eq{HSOL} provided all $T_k$ are real. The goal of this appendix is to show that once the following equations (see \eqref{eq:RHOu}) are
satisfied
\beqa
\la{eq:1}
\frac{T_0^{u+}}{\Phi_u}&=&\frac{\frac{g^{+}}{\bar g^+}\Q^{++}_u \Q_v^{++}- \bar \Q_u\bar\Q_v}{\frac{g^{+}}{g^-}\Q_u^{++}\bar \Q_v^{--}-\Q_u\bar\Q_v}\;,
\\ \la{eq:2}
-\frac{T_0^{v+}}{\bar\Phi_{v}}&=&\frac{\frac{ g^+}{\bar g^+} \Q_v^{{++}}
   \Q_u^{{++}}-\bar\Q_v \bar \Q_u }{\frac{\bar  g^-}{\bar g^+}\Q_v^{{++}}\bar{\Q}_u^{{--}}
   - \Q_v \bar{\Q}_u  }\;,\\ \la{eq:3}
   \Phi_v&=&g^+g^-\Phi_u\;,
\eeqa
and $T_0^{u}$ and $T_0^v$ are real then all $T_k$ are real and thus all the formulae in the main text go through and the $Y$-system is indeed solved by (\ref{TY}). Before proving this statement we recall that
\eq{eq:gbg} follows directly from \eq{eq:3}
under certain analyticity assumptions and also from
$\frac{T_0^{u+}T_0^{u-}}{\Phi_u\bar\Phi_u}=\frac{T_0^{v+}T_0^{v-}}{\Phi_v\bar\Phi_v}$,
which is a consequence of \eq{eq:1} and \eq{eq:2}. Thus we can add the equation
\beq
T_0^v=g\bar
gT_0^u \la{new}
\eeq
to the equations at hand and proceed to the proof of the reality of the $T$-functions.

Equation (\ref{eq:1}) implies
\beq\la{eq:h1}
\Phi_{u}=ih_1^+\left(\frac{\Q_u^{++}\bar \Q_v^{--}}{g^-}-\frac{\Q_u \bar \Q_v}{g^+}\right)\;\;,\;\; T_0^{u}=ih_1\left(\frac{\Q_u^{+}\Q_v^{+}}{\bar g}-\frac{\bar \Q_u^{-} \bar
\Q_v^{-}}{g} \right)\;,
\eeq
for some $h_1$. Since $T_0^u$ is real for real $x$ the function $h_1$ is a real function. \Eq{eq:2}
implies
\beq
\bar\Phi_{v}=-ih_2^+\left(\bar g^{-}\Q_v^{++}\bar \Q_u^{--}-\bar g^+\Q_v \bar \Q_u\right)\;\;,\;\; T_0^{v}=ih_2\left(g\Q_u^{+}\Q_v^{+}-\bar g\bar \Q_u^{-} \bar
\Q_v^{-} \right), \la{here2}
\eeq
where again $h_2$ is a real function. In virtue of (\ref{new}) we have $h_1=h_2\equiv h$ and
by conjugating the first equation in (\ref{here2}) we find
\beq
\Phi_{v}=ih^-\left(g^{+}\bar\Q_v^{--} \Q_u^{++}-g^-\bar\Q_v  \Q_u\right)
=\frac{h^-}{h^+}g^-g^+\Phi_u\;,
\eeq
which means that the real function $h(x)$ is periodic in the imaginary direction,
$h(x)=h(x+i)$. This in turn implies that the function $h^-$ is also a real function because
$$\overline{
h(x-i/2)}=h(x+i/2)=h(x-i/2)\,.$$
Then it is simple to see that $T_1^u$ is real. We simply write, from \eq{eq:h1},
\beq
T_1^u=T_0^{u-}\frac{\Q^{++}_u}{\Q_u}+\Phi_u\frac{\bar \Q^{--}_u}{\Q_u}=i h^-\left(\frac{\Q_v\Q_u^{++}}{\bar g^-}-\frac{\bar\Q_v\bar\Q_u^{--}}{g^+}\right)\;.
\eeq
The reality of $T_1^u$ is now manifest because
the expression inside the brackets is purely imaginary and, as we have just shown, $h^-$ is real. Proceeding in the same way one can see that all $T_k^u$ are real and thus
the Hirota equation is satisfied by our solution.

\section{Proof of analyticity of $T_{-1}$ in the physical strip}
\label{ANALYT}
\subsection{Analyticity in the $U(1)$ sector}
From (\ref{eqF}) together with $F(x\pm i/2 \pm i0)=F(x\pm i/2 \mp i0)-T_{-1}(x)$ we have
%\beq
%T_{-1}(x)=(F(x+i/2)+F(x-i/2)) \frac{F(x+i/2+i0)F(x-i/2-i0)}{
%\[F(x+i)F(x-i)\]^{*2s}}\,e^{-L\cosh(\pi x)}\;, \nn
%\eeq
%\beqa
%T_{-1}(x)&=&(F(x+i/2+i0)+F(x-i/2+i0)) \frac{F(x+i/2+i0)\left(F(x-i/2+i0)-T_{-1}(x)\right)}{
%\[F(x+i)F(x-i)\]^{*2s}}\,e^{-L\cosh(\pi x)}\nn\\
%&=&(F(x+i/2-i0)+F(x-i/2-i0)) \frac{\left(F(x+i/2-i0)-T_{-1}(x)\right) F(x-i/2-i0)}{
%\[F(x+i)F(x-i)\]^{*2s}}\,e^{-L\cosh(\pi x)}\nn\, .
%\eeqa
%which can be solved for $T_{-1}$ to yield
\beq
T_{-1}(x)=\frac{T_1 F^{+_{\!\!+}} F^{-_{\!\!+}}\,}{\[F^{++}F^{--}\]^{*2s}e^{L\cosh(\pi x)}+T_1 F^{+_{\!\!+}}\,}
=\frac{ T_1F^{+_{\!\!-}} F^{-_{\!\!-}}\,}{\[F^{++}F^{--}\]^{*2s}e^{L\cosh(\pi x)}+T_1 F^{\MM}\,}\;.
\eeq
Using respectively the first/second equality we can smoothly move from real $x$ into the upper/lower half complex plane provided $T_1$ is analytic. In this way we can reach any $x$ inside the enlarged strip $|\im(x)|<1$ where $T_1$ is regular.

Notice that for large $L$ we have
\beqa
T_{-1}(x)& \simeq & \left\{ \begin{array}{ll}
\phi^{-} & ,   +1/2<\im(x)<+1\\
 0 &  ,  -1/2<\im(x)<+1/2\\
\phi^{+} &   ,-1\,\,\,\,\,\,<\im(x)<-1/2\end{array}\right.
\eeqa
The denominator in the expression for $T_{-1}$ at $x=\theta_j-i/2$  is proportional to Bethe equations $ S^2 e^{ip_jL}+1=0 $. This is not a pole of $T_{-1}$ because the numerator at these points is proportional to $T_0(\theta_j)=0$.

However, for large volume, $T_{-1}$ could have poles at the analogue of the holes of the $\theta_j$ BAE, close to the boundaries of the physical strip.

\subsection{General case}
In this subsection we will study the analyticity of $T^{u}_{-1}(x)$ and $T^v_1(x)$ for a general solution. We will show that for large enough $L$ these functions are analytic inside the physical strip $-1/2+\epsilon<\IM x<1/2-\epsilon$ where $\epsilon\to 0$ when $L\to\infty$.
We start from \eq{Tgen} and rewrite it as
\beq
\la{Tgen2}
%T_{+1}^v &=& %e^{-L\cosh(\pi x)}\left(\frac{\Phi_v^+\bar\Phi_v^-}{\Phi_u^+\bar\Phi_u^-}\right)^{*s}T_{+1}^u=
%-e^{-L\cosh(\pi x)}
% \frac{\left(T_0^{v-} \Q_v-T_{+1}^v\Q_v^{++}\right)\Phi_v T_{+1}^u}{\bar %\Q_v(\Phi_u^+\Phi_v^-\bar\Phi_v^+\bar\Phi_u^-)^{*s}}\nn \,,\\
T_{-1}^u = +e^{-L\cosh(\pi x)} \frac{\left(T_0^{u-} \Q_u-T_{-1}^u\Q_u^{++}\right)\bar\Phi_u
T_{-1}^v}{\bar \Q_u(\Phi_u^+\Phi_v^-\bar\Phi_v^+\bar\Phi_u^-)^{*s}} \,.
\eeq
and similar for $T_{+1}^v$. Solving for \(T^u_{-1}\) we  get
\beq
T_{-1}^u=\frac{e^{-L\cosh(\pi x)}T_0^{u-}\Q_u}{\bar Q_u(\Phi_u^+\Phi_v^-\bar\Phi_v^+\bar\Phi_u^-)^{*s}-e^{-L\cosh(\pi x)}\bar\Phi_u T_{-1}^{v}\Q_u^{++}}
\eeq
Since $\bar\Phi_u$ is regular in the lower half plane and $T_{-1}^v$ is regular
in the  strip $-1<\IM x<1$ where $(\Phi_u^+\Phi_v^-\bar\Phi_v^+\bar\Phi_u^-)^{*s}$
is also regular the singularities of $T_{-1}^u$ for $-1<\IM x<0$ could be only due to zeros
of the denominator.

As far as $L$ or $x$ are large for $-1/2<\IM x<0$, the second term in the
denominator is exponentially suppressed and to get a zero of the denominator we should be close to a zero of $\bar Q_u$. The points close to $\bar u_j$ where the denominator
vanishes are in fact $x=u_j$ as follows from the auxiliary finite volume BAE
\eq{baeu0}. However these poles cancel with zeroes of the numerator rendering $T_{-1}$ regular at these points, a result we were familiar with already. Thus, we see that
for large $L$ the only poles that could appear must lie close to the border of the physical
strip $\IM x=-1/2 $ where the exponent is oscillating. It oscillates faster
for large $x$ and we thus have poles condensing at infinity along the
borders of the physical strip.

We conclude that for the general solution -- at least for large $L$'s -- the function $T_{-1}^u(x)$
is analytic inside almost the whole physical strip and could have poles only
very close to the border.
%However, for large volume, $T_{-1}$ could have poles at the analogue of the holes of the $\theta_j$ BAE, close to the boundaries of the physical strip.
For small $L$ it can probably happen that the singularities approach the real axe. That could indicate some singular behavior of the energy levels as functions of $L$ such as the one observed in \cite{Dorey:2000rv} (see e.g. figure 10 in this work). It would be interesting to investigate these points in greater detail.

\section{Details on L\"uscher formulae derivation}
\label{appE}
If this section we shall present some details of the computation of the first finite volume correction to the asymptotic auxiliary Bethe equations, obtained by expanding (\ref{baeu0}) to the leading order (see section \ref{Luscher:sec} for notation). We start by writing
\beq
P(U_j)=\exp\[K_1(U_j-i/2+i0)*\rho_u-K_1(U_j+i/2+i0)*\bar \rho_u \] \,.
\eeq
Notice that we introduced the $i0$'s because $P(x)$ was originally defined in (\ref{Px}) for $x$ in the upper-half-plane. Removing the $i0$'s by the use the Sokhatsky-Weierstrass formula we get
\beq
P(U_j)=\exp\[K_1^-*\rho_u-K_1^+*\bar \rho_u+Y_0/2\]_{x=U_j} \,,
\eeq
where the convolutions  are understood in the principal part sense and $\rho_u+\bar \rho_u=\log(1+Y_0)\simeq Y_0$ was used. Next we split the density $\rho_u$ into $\rho^{(1)}_u$ and $\rho_u^{(2)}$ as explained in section (\ref{Luscher:sec}). The former contribution is purely imaginary and therefore it contributes to the exponent as
\beq
K_1^-*\rho_u^{(1)}-K_1^+*(- \rho_u^{(1)})=K_2*\rho_u^{(1)} \,.
\eeq
Hence we finally obtain
\beq
P(U_j)=\exp\left( K_{2}*\rho^{(1)}+(K_1^{-}*\rho^{(2)}-c.c.)+\frac{Y_0}{2}\right)_{x=U_j} \,,
\eeq
with all convolutions understood as principal part integrals. It turns out that the first and the last terms in this exponent simply convert the Bethe roots $u_j$ in (\ref{baeu0}) into their real parts, namely,
\beqa
\frac{ \Q_u^{++}(u_i)}{\bar \Q_u^{--}(u_i)}\exp\left(K_2*\rho^{(1)}(u_i)\right)
&=&\frac{ \QQ_u^{++}(u_i)}{ \QQ_u^{--}(u_i)}\\
 \frac{\phi^-(u_i)}{\phi^+(u_i)}\frac{ \QQ_u^{++}(u_i)}{\QQ_u^{--}(u_i)}\exp\left(\frac{Y_0(U_i)}{2}\right)
 &=&
  \frac{\phi^-(U_i)}{\phi^+(U_i)}\frac{ \QQ_u^{++}(U_i)}{\QQ_u^{--}(U_i)}\;.
\eeqa
The check of the first equality goes exactly as in (\ref{derivation}) and we will therefore not consider it here. Let us  explain how to check the second equality. Notice that this expression is equivalent to
\beqa
\Delta u_j \, \partial_x \log\left( \frac{\phi^-(x)}{\phi^+(x)}\frac{ \Q_u^{++}(x)}{\Q_u^{--}(x)} \right)_{x=U_j} = -\frac{Y_0(U_j)}{2} \la{Du} \,.
\eeqa
Next, we write the right hand side containing $Y_0\simeq T_{-1}T_1/\phi^+\phi^-$ as
\beqa
-\frac{Y_0(U_j)}{2}=\left(-\frac{\Q^{++}_u (U_j)T_{-1}^u(U_j)}{2\phi^+(U_j) \Q'_u(U_j)} \right) \left( \frac{T_1^u(U_j) \Q'_u(U_j) }{\Q_u^{++}(U_j) \phi^-(U_j)} \right) \,. \la{parent}
\eeqa
The first factor in the r.h.s. is precisely $\Delta u_j$. This can be seen from expanding the second equation from (\ref{JUMPS}) at $x=\bar u_j=u_j-2 \Delta u_j$ to leading order in $\Delta u_j$. Alternatively we can find the imaginary part of $u_j$ by imposing regularity on the density (\ref{eq:RhouCorr}) at $x=u_j$.  To simplify the second factor in (\ref{parent}) we write
\beq
T^u_1(U_j)=\left.\frac{\left(\Q_u^{++}\phi^- + \Q_u^{--}\phi^+  \right)'}{\Q_u'} \right|_{x=U_j}
\eeq
Evaluating the derivative of the numerator and using the leading order auxiliary Bethe equations $\Q^{--}_u(U_j)\phi^+(U_j)+ \Q_u^{++}(U_j)\phi^-(U_j)\simeq 0$ we find
\beq
 \frac{T_1^u(U_j) \Q'_u(U_j) }{\Q_u^{++}(U_j) \phi^-(U_j)} = \partial_x \log\left( \frac{\phi^-(x)}{\phi^+(x)}\frac{ \Q_u^{++}(x)}{\Q_u^{--}(x)} \right)_{x=U_j}
\eeq
thus identifying the second factor in the left hand side of (\ref{Du}) and completing our proof. Therefore the expansion of the auxiliary Bethe equation (\ref{baeu0}) simply reduces to (\ref{corrU}), as announced in the main text.

To simplify the Bethe equations (\ref{corrTh}) and (\ref{corrU}) further we shall relate the convolutions in these expressions to particular derivatives of the $Y$-function $Y_0$. To compute these derivatives it is useful to notice that we can write $Y_0$ in terms of two simple pure phase functions $a_u$ and $a_v$,
\beq
a_w(x)=S(x) \frac{\Q^+_w(x)}{\Q^-_w(x)} e^{iL/2 \sinh(\pi x)}
\eeq
as
\beq
Y_0=\left(a_u^++1/a_u^-\right)\left(a_v^++1/a_v^-\right)  \,.
\eeq
In this form, it is easy to compute the derivative of $Y_0$ with respect to $\theta_k$, $u_k$ or $v_k$ because we can use a simple identity
\beq
\partial_{\theta_i}a_w(x)=-\pi i K_0(\theta_i-x)a_w(x)\;\;,\;\;
\partial_{w_i}a_w(x)=2\pi i K_1(w_i-x)a_w(x)\;.
\eeq
%where $K_k(x)\equiv \frac{1}{2\pi i }\partial_x \log S_0^2(x-\theta_k)$.
Furthermore if we notice that the densities $\rho^{(2)}$ can also be simply expressed in terms of these new functions as
\beq
\rho^{(2)}_u=a_u^+\left(a_v^++1/a_v^-\right) \,\, , \,\, \rho^{(2)}_v=a_v^+\left(a_u^++1/a_u^-\right)\;,
\eeq
It is then a straightforward exercise to check the identities (\ref{remark}).

%%%%%%%%%%%%%%%%%%%%%%%%%%%%%%%%%%%%%%%%%

\printindex

\end{document}